\documentclass[aps,nofootinbib,amsmath,prd,twocolumn,showpacs,superscriptaddress,groupedaddress]{revtex4-1}

\usepackage[american]{babel}
\usepackage{amsfonts}
\usepackage{amsmath}
\usepackage{amssymb}
\usepackage{slashed}
\usepackage{xcolor}
\usepackage{bm}
\usepackage{float}
\usepackage[utf8]{inputenc}
\usepackage[macrosonly]{chet}
\usepackage{nicefrac}
\usepackage{hyperref}
\usepackage{url}

\begin{document}

\title{Consequences of the gauging of Weyl symmetry
\\
and the two-dimensional conformal anomaly}

\author{Omar Zanusso}
\email{omar.zanusso@unipi.it}
\affiliation{Universit\`a di Pisa and INFN - Sezione di Pisa, Largo Bruno Pontecorvo 3, 56127 Pisa, Italy}

\begin{abstract}
%
We discuss the generalization of the local renormalization group approach 
to theories in which Weyl symmetry is gauged. These theories naturally correspond to scale invariant -- rather than conformal invariant -- models in the flat space limit.
We argue that this generalization can be of use when discussing the issue of scale vs conformal invariance in quantum and statistical field theories.
The application of Wess-Zumino consistency conditions constrains the form of the Weyl anomaly and the beta functions in a nonperturbative way. In this work we concentrate on two dimensional models including also the contributions of the boundary. Our findings suggest that the renormalization group flow between
scale invariant theories differs from the one between conformal theories because of the presence of a new charge that appears in the anomaly.
It does not seem to be possible to find a general scheme for which the new charge is zero, unless the theory is conformal in flat space.
Two illustrative examples involving flat space's conformal and scale invariant models that do not allow for a naive application of the standard local treatment are given.
\end{abstract}

\maketitle

\section{Introduction}\label{sect:intro}

In essence, the local renormalization group (rg) is a generalization of the standard renormalization group to local couplings in curved space \cite{Shore:1986hk}. The local approach has the advantage that the position-dependent couplings and the metric act as sources for the expectation values of the interaction operators and the energy-momentum tensor,
assuming of course the existence of a finite and renormalized effective action from the path-integral.

One of the main achievements of the local rg comes by combining it with the analysis of the Wess-Zumino consistency conditions \cite{Wess:1971yu} to Weyl transformations \cite{Osborn:1991gm}. By simply requiring that local scale transformations are Abelian, it is in fact possible to rederive the famous Zamolodchikov's result of irreversibility of the rg in two dimensions (the famous $C$-theorem) \cite{Zamolodchikov:1986gt}
in an elegant way by constraining the flow of local charges of the Weyl anomaly and the beta functions of marginal couplings \cite{Osborn:1991gm}.
It is also possible to generalize some of the same results to four dimensions \cite{Jack:1990eb}, resulting in a perturbative proof of irreversibility,
i.e., a perturbative $A$-theorem \cite{Cardy:1988cwa}.

In this paper we generalize the two-dimensional local rg approach
to the case in which Weyl symmetry is realized as a gauge symmetry, so there is an additional vector gauge potential $S_\mu$
which has an affine behaviour under Weyl transformations, $S_\mu \to S_\mu -\partial_\mu \sigma$. The new potential
is a source for the dilation current \cite{Iorio:1996ad}, which we denote as $D^\mu$. Classically the gauge symmetry implies that the trace of the energy momentum tensor is related to the divergence of $D^\mu$ as
\begin{equation}
 \begin{split}
  T^\mu{}_\mu = \nabla_\mu D^\mu
  \,,
 \end{split}
\end{equation}
implying that the gauged theory is \emph{scale} invariant, rather than conformal invariant, in the flat space limit.
While the standard local rg approach is particularly useful for discussing the properties of the renormalization flow between two conformal field theories (CFTs), we deduce that the gauged counterpart considered in this paper
is relevant for theories that are simply scale invariant, given the natural identification that can be made in the limit of flat space between the dilation current $D_\mu$ and the virial current that characterizes scale invariant theories \cite{Nakayama:2013is}.

The Wess-Zumino consistency of the gauged Weyl group
is structurally interesting because dilatations do not commute with the local Lorentz group in curved space, so the gauge-covariant derivative includes a departure from the standard Christoffel components that depends on $S_\mu$ \cite{Codello:2012sn}, which is known as a special type of ``disformation'' in the metric-affine literature \cite{Sauro:2022chz}.
This results in a ``mixture'' between the structures of the standard local rg and those of the application of Wess-Zumino conditions to an Abelian gauge theory.
The most important result of this paper is that in two dimensions it is possible to
find a quantity, denoted $\tilde{\beta}_\Psi$, which on renormalization group trajectories behaves as
\begin{equation}
 \begin{split}
  \mu\frac{{\rm d}}{{\rm d}\mu}{\tilde{\beta}}_\Psi
  =
  \chi_{ij}\beta^i\beta^j +\beta^S_2
  \,,
 \end{split}
\end{equation}
where $\mu$ is the renormalization group scale, $\chi_{ij}$ can be interpreted as a metric in the space of couplings, $\beta^i$ are the beta functions of the marginal couplings, and $\beta^S_2$ is an extra contribution associated to a new charge that appears in the Weyl anomaly only for gauged symmetry. The quantity $\tilde{\beta}_\Psi$ generalizes $\tilde{\beta}_\Phi$ considered by Osborn to rederive Zamolodchikov's theorem \cite{Osborn:1991gm}. However, irreversibility of the renormalization group flow is not possible for an arbitrary value of $\beta^S_2$, unless there exists a scheme for which $\beta^S_2$ is zero, which is not true in general, and the metric $\chi_{ij}$ is positive definite, which is guaranteed for unitary theories. Using two nonunitary theories as examples, we confirm that $\beta^S_2$ is in general nonzero, and we show that it is zero in two specific example of interest for CFT (including a special case of the theory of elasticity).

As for the organization of the paper,
we discuss the symmetries of interest in the remaining parts of Sect.~\ref{sect:intro}, both from the classical and quantum points of view.
Sect.~\ref{sect:two} contains the aforementioned main result for the consistency of the anomaly, including geometric conditions on how to set $\beta^S_2=0$ consistently on the flow of models that do allow it, and on the structure of the boundary terms.
Sect.~\ref{sect:examples} shows the application of the result to two examples, including a special higher derivatives CFT which does not allow the standard local rg treatment, and the theory of elasticity, which is known to be scale but not conformal invariant in flat space.
Sect.~\ref{sect:interplay} discusses qualitatively the interplay of the anomaly with diffeomorphism invariance and some consequence.
A brief comparison between Riemannian and Weylian geometries is reported in Appendix~\ref{sect:geometry}, while some details on the examples' computations of the anomalies are given in Appendix~\ref{sect:hk}.

Throughout the paper we adopt the Euclidean convention and discuss everything in terms of the effective action $\Gamma$ rather than the generator of connected correlators $W$. Furthermore, all formulas can be generalized to Lorentzian signature with very little additional work.

\subsection{Classical conformal and scale vs gauged Weyl symmetries}\label{sect:classical}

Consider a \emph{classical} action $S[\Phi,g]$ of some field $\Phi$
on a Riemannian manifold with metric $g_{\mu\nu}$.
Standard Weyl transformations are defined as
\begin{equation}\label{eq:weyl-transformations}
 \begin{split}
  g_{\mu\nu} \to g'_{\mu\nu} = {\rm e}^{2\sigma} g_{\mu\nu}\,,
  \qquad
  \Phi \to \Phi'= {\rm e}^{w_\Phi \sigma} \Phi\,,
 \end{split}
\end{equation}
where metric and field are rescaled by a local positive function, $\sigma=\sigma(x)$, according to their Weyl weights $w(g_{\mu\nu})=2$ and $w_\Phi$. Invariance of $S[\Phi,g]$ under conformal transformations implies that
the variational energy-momentum tensor, $T^{\mu\nu}=-\frac{2}{\sqrt{g}}\frac{\delta S}{\delta g_{\mu\nu}}$ is traceless, $T=g_{\mu\nu}T^{\mu\nu}=0$ when $\Phi$ is on-shell, i.e., when $\frac{\delta S}{\delta \Phi}=0$. Assuming also invariance under diffeomorphisms, we have that Weyl symmetry implies flat space's conformal symmetry in the limit
$g_{\mu\nu}\to \delta_{\mu\nu}$, and therefore Weyl symmetry can be regarded as the natural extension of conformal symmetry to curved space \cite{Brown:1980qq}.\footnote{%
More precisely, the group of conformal isometries which leave a given metric $g_{\mu\nu}$ invariant is a subgroup of the semidirect product of the diffeomorphisms and Weyl groups, which is finite in $d>2$. So the flat space limit of a Weyl invariant theory must be a CFT and we can construct currents with the energy-momentum tensor.
}

Constant scale transformations are the subgroup of Weyl transformations such that $\sigma$ is constant and, as such, they constrain much less the trace of energy-momentum tensor. In particular, we have that the integral of $T$ is zero,
$\int \sqrt{g} T=0$, implying that $T=\nabla^\mu J_\mu$, for some vector $J_\mu$ known as the virial current \cite{Nakayama:2013is}. If the virial current satisfies some properties, for example if it is the divergence of a symmetric tensor $J_\mu=\nabla^\nu X_{\mu\nu}$, we have that a scale invariance action can be ``improved'' to a Weyl-invariant one by including off-shell couplings
among the curvatures and the tensor $X_{\mu\nu}$ in $d>2$ \cite{OsbornLectures}.
A stronger condition is needed in $d=2$ to have full Virasoro invariance \cite{Polchinski:1987dy}.

Weyl symmetry is a local symmetry, but it is not a gauge symmetry in the traditional sense \cite{Iorio:1996ad}. In fact, it is not a straightforward task to write down a Weyl-symmetric action for an arbitrary field, conversely
it is generally simple to write a scale invariant one by just applying dimensional analysis.
The main reason of the difficulty is that the Levi-Civita covariant derivative $\nabla_\mu$ does not transform covariantly under Weyl transformations, because it depends on the derivatives of the metric $g_{\mu\nu}$, so spurious terms in the transformation of $\nabla_\mu \Phi$
must be cancelled by opportune couplings with the curvatures. The construction of Weyl-covariant connections requires in general additional geometrical structures \cite{Fefferman:2007rka}.

There is, however, a gauged version of Weyl symmetry, which we refer to as gauged Weyl symmetry. It includes a vector potential $S_\mu$ that has an affine transformation \cite{Weyl:1918ib} so that \eqref{eq:weyl-transformations} are replaced by
\begin{equation}\label{eq:gauged-weyl-transformations}
 \begin{split}
  & g_{\mu\nu} \to g'_{\mu\nu} = {\rm e}^{2\sigma} g_{\mu\nu}\,,
  \qquad
  S_\mu \to S'_\mu = S_\mu -\partial_\mu \sigma\,,\\
  &
  \Phi \to \Phi'= {\rm e}^{w_\Phi \sigma} \Phi\,,
 \end{split}
\end{equation}
and there also exists a gauged Weyl covariant derivative $\hat{\nabla}$, which we show below.
In a classical action $S=S[\Phi, g, S_\mu]$ (we use the same symbol for the action and the gauge potential, hopefully it does not generate confusion since the latter always has an index),
the field $\Phi$ couples to the new vector through the dilation current, $D^\mu = \frac{1}{\sqrt{g}}\frac{\delta S}{\delta S_\mu}$. The main consequence of gauged Weyl symmetry is that the trace of the energy-momentum tensor equals the divergence of the dilation current,
$T=\nabla^\mu D_\mu$. This obviously implies that, in the flat space limit $g_{\mu\nu}\to \delta_{\mu\nu}$ and $S_\mu \to 0$,
gauged Weyl-invariance reduces scale invariance \cite{Iorio:1996ad} and also suggests the identification of $D_\mu$ with the virial current in the same limit \cite{Sauro:2022chz}.

In order to see that gauged Weyl invariance is actually the natural generalization of scale invariance to curved space, it is important to understand the underlying geometry. Using the gauge potential $S_\mu$, we can construct a covariant derivative which is covariant both under diffeomorphisms and gauged Weyl transformations \cite{Codello:2012sn,Sauro:2022chz},
\begin{equation}\label{eq:nabla-hat}
 \begin{split}
  \hat{\nabla}_\mu \Phi = \nabla_\mu \Phi + L_\mu \cdot \Phi + w_\Phi S_\mu \Phi\,.
 \end{split}
\end{equation}
The new connection consists of the Levi-Civita symmetric term such that $\nabla_\mu g_{\nu\rho}=0$, a new term with components $(L_\mu)^\alpha{}_\beta = L^\alpha{}_{\beta\mu} = \frac{1}{2}(S_\beta \delta^\alpha_\mu +S_\mu \delta^\alpha_\beta- S^\alpha g_{\beta\mu})$,
and a multiplicative (gauge) term weighed by the charge $w_\Phi$.
The new connection is Abelian, like a standard Maxwell potential, but,
structurally, differs from Maxwell's because of the contribution $L_\mu$, known in the metric-affine literature as a special type of disformation.
In fact, the disformation encodes the fact that
local dilatations and Lorentz transformations do not commute \cite{Sauro:2022hoh,Karananas:2015eha}.
The contributions are seen as coming from the noncommutativity of the generators of the subgroup $D_1 \ltimes SO(d)$ with the rest of $GL(d)$ \cite{Tomboulis:2011qh}, where $D_1$ are the gauged Abelian dilatations.

The gauged connection is also compatible with the metric,
$\hat{\nabla}_\mu g_{\nu\rho}=0$, because $w(g_{\mu\nu})=2$.
It can also be integrated by parts using the density $\sqrt{g}$, i.e.,
$ \sqrt{g} \hat{\nabla}_\mu v^\mu = \partial_\mu (\sqrt{g} v^\mu)
$
iff the weight of the vector satisfies $w_v = -d$, i.e., it is a Weylian density as well \cite{Sauro:2022chz}. Most importantly, 
the derivative of a field transforms like the field itself
\begin{equation}
 \begin{split}
  \hat{\nabla}_\mu \Phi \to \hat{\nabla}'_\mu\Phi'= {\rm e}^{w_\Phi \sigma} \hat{\nabla}_\mu \Phi\,,
 \end{split}
\end{equation}
thanks to the fact that the transformation of $S_\mu$ cancels precisely the noncovariant contributions coming from the Christoffel connection.
For these reasons, any scale invariant action can immediately be promoted to a gauged Weyl invariant action through the replacements of $\nabla_\mu \to \hat{\nabla}_\mu$ and of Riemannian curvatures with curvatures of $\hat{\nabla}$ \cite{Ghilencea:2023wwf}. There is, in addition,
a new gauge-invariant field-strength $W_{\mu\nu}= \partial_\mu S_\nu-\partial_\nu S_\mu=2 \partial_{[\mu}S_{\nu]}$ \cite{Sauro:2022chz}, similar to Maxwell's.
With all the above properties, it becomes trivial to write down Weyl-invariant actions, for example
$
\int \sqrt{g} \hat{\nabla}_\mu \varphi \hat{\nabla}^\mu \varphi
$
is manifestly invariant in any dimension $d$ for a scalar field $\varphi$ that weighs $w_\varphi=\frac{2-d}{2}$. More details on the geometry of gauged Weyl symmetry and spacetime are given in Appendix~\ref{sect:geometry}.

\subsection{Quantum symmetry and Wess-Zumino consistency}\label{sect:quantum}

It is well-known that quantum symmetries can be anomalous for an effective action $\Gamma$ coming from a path-integral construction.
Schematically, we have
\begin{equation}\label{eq:pi}
 \begin{split}
  {\rm e}^{-\Gamma} = \int [{\rm d}\Phi] \, {\rm e}^{-S}\,,
 \end{split}
\end{equation}
where at the exponent on the lhs we have the finite and renormalized generating functional of connected Green functions and $\Phi$ is some bare field over which we are integrating.
The functional $\Gamma$ is assumed to be a finite functional of the couplings $\lambda^i$, the metric $g_{\mu\nu}$, and the gauge potential $S_\mu$, i.e., $\Gamma=\Gamma[\lambda^i,g_{\mu\nu},S_\mu]$. It satisfies a Callan-Symanzik equation
\begin{equation}\label{eq:cs-eq}
 \begin{split}
  \left(\mu \frac{\partial}{\partial \mu} + \beta^i \frac{\partial}{\partial \lambda^i}\right) \Gamma=0\,,
 \end{split}
\end{equation}
which defines the beta functions $\beta^i = \mu \frac{{\rm d}}{{\rm d}\mu} \lambda^i$ with rg scale $\mu$. We assume a dimensionless regularization scheme, such as dimensional regularization, and concentrate on marginal couplings $\lambda^i$.

Symmetries involving transformations of the local scale are naturally anomalous because of the presence of rg beta functions. However, conformal symmetry has intrinsic anomalies that are most easily seen by considering the theory in curved space \cite{Brown:1980qq,Polchinski:1987dy}. We want to generalize the analysis based on local rg by Osborn \cite{Osborn:1991gm} to gauged Weyl symmetry, so
we assume that the path-integral \eqref{eq:pi} can be consistently renormalized for \emph{local} couplings $\lambda^i=\lambda^i(x)$.
In general, the procedure
requires additional counterterms involving derivatives of the coupling themselves \cite{Osborn:1991gm}, and the specific form of the additional counterterms in some examples will become clearer below.
The metric, the gauge potential, and the local couplings are sources for the bare energy-momentum tensor, dilation current and interaction operators, respectively.
By construction, derivatives of $\Gamma$ give their expectation values
\begin{equation}\label{eq:expectation-values}
 \begin{split}
  &\langle T^{\mu\nu}\rangle = -\frac{2}{\sqrt{g}}\frac{\delta \Gamma}{\delta g_{\mu\nu}} \,,\qquad 
  \langle D^\mu\rangle = \frac{1}{\sqrt{g}}\frac{\delta \Gamma}{\delta S_\mu} \,,\\
  &
  \langle {\cal O}_i\rangle = -\frac{1}{\sqrt{g}}\frac{\delta \Gamma}{\delta \lambda^i}
  \,,
 \end{split}
\end{equation}
where ${\cal O}_i$ are the interaction operators associated to the couplings $\lambda^i$.

To begin with, we slightly generalize the construction of Ref.~\cite{Osborn:1991gm} and define the operator
\begin{equation}\label{eq:weyl-delta}
 \begin{split}
  \Delta^W_\sigma = \int \Bigl\{2\sigma g_{\mu\nu} \frac{\delta}{\delta g_{\mu\nu}}
  -\partial_\mu \sigma\frac{\delta}{\delta S_\mu}
  \Bigr\}\,.
 \end{split}
\end{equation}
It should be obvious that $\Delta^W_\sigma S=0$ on-shell for an invariant \emph{classical} action $S$ with constant couplings, as discussed in the previous section. In fact it straightforwardly implies the classical relation $T=\nabla_\mu D^\mu$.
In general, for the \emph{quantum} action $\Gamma$ of \eqref{eq:pi}, we must have
\begin{equation}
 \begin{split}
  \Delta^W_\sigma \Gamma = \Delta^\beta_\sigma \Gamma + A_\sigma\,,
 \end{split}
\end{equation}
where the first term accounts for scale-dependence caused by the renormalization's beta functions and the second term is the ``true'' anomaly caused by the curved geometry as well as the additional counterterms to the local couplings. The above relation can be rearranged as
\begin{equation}\label{eq:full-delta}
 \begin{split}
\Delta_\sigma \Gamma \equiv(\Delta^W_\sigma -\Delta^\beta_\sigma)\Gamma =  A_\sigma\,,
 \end{split}
\end{equation}
and the transformation $\Delta_\sigma=\Delta^W_\sigma -\Delta^\beta_\sigma$ now accounts for all contributions to the change of the local scale, both classical and quantum \cite{Fortin:2012hn}.
Since the original transformation is Abelian, it must be that also its quantum counterpart $\Delta_\sigma$ remains Abelian, even if it is anomalous because of the presence of $A_\sigma$.
Therefore we can impose the Wess-Zumino consistency condition \cite{Wess:1971yu}
\begin{equation}\label{eq:wz-consistency-general}
 \begin{split}
  [\Delta_\sigma,\Delta_{\sigma'}] \Gamma = 0\,,
 \end{split}
\end{equation}
which enforces the Abelian nature of $\Delta_\sigma$
and translates into a consistency condition for the anomaly
\begin{equation}
 \begin{split}
  (\Delta^W_\sigma -\Delta^\beta_\sigma) A_{\sigma'}- (\sigma \leftrightarrow \sigma')=0\,.
 \end{split}
\end{equation}
The consistency condition constrains the local structures of the integrand
of the anomaly with the beta functions of the renormalized action \cite{Osborn:1991gm}.

For example, in the case of only marginal couplings $\lambda^i$ to local operators ${\cal O}_i$, for the procedure to work consistently then the couplings $\lambda^i$ \emph{must} be local functions \cite{Fortin:2012hn}
and $\Gamma \supset -\int \sqrt{g} \lambda^i \langle{\cal O}_i \rangle $.
In this case we have
\begin{equation}\label{eq:beta-delta}
 \begin{split}
  \Delta^\beta_\sigma = -\int \sigma \beta^i \frac{\delta}{\delta \lambda^i}
  \,,
 \end{split}
\end{equation}
which includes the beta functions $\beta^i$ of the marginal couplings $\lambda^i$. 
The beta functions are functions of $\lambda^i$
and can be thought of as the components of a vector in the space of couplings.
For consistency of the local approach, we must have that $A_\sigma$ is a local functional of $g_{\mu\nu}$ and
$S_\mu$, and also of covariant tensors constructed from $\lambda^i$ and their derivatives.
Essentially, this was tacitly assumed when declaring renormalizability \cite{Osborn:1991gm}.
The number of derivatives in $A_\sigma$ depends on the dimensionality of spacetime and the case $d=2$ is worked out in the next section under the assumption of a dimensionless rg scheme.
For a much more thorough discussion on how to pass from a standard to a local rg scheme and a comprehensive view on the topic we refer to Ref.~\cite[Sect.~II]{Fortin:2012hn} (cautioning that we use slightly different conventions).

\section{Consistency conditions in two dimensions}\label{sect:two}

In $d=2$ the curvature scalar of $\hat{\nabla}$ becomes $\hat{R}=R-2 \nabla^\mu S_\mu$, where $R=R[g]$ is the curvature scalar of the Levi-Civita connection $\nabla$ and depends only on the metric (see the discussion of Appendix~\ref{sect:geometry}), so it is a natural term to parametrize the anomaly, because,
for a dimensionless regularization scheme, we must have that the anomaly $A_\sigma$ is the integral of local dimension two operators constructed with the available sources \cite{Osborn:1991gm}.\footnote{%
Our approach based on a dimensionless scheme differs from the one of Ref.~\cite{Morris:2018zgy}
which uses an infrared cutoff.
} 
We choose to parametrize the anomaly as
\begin{equation}\label{eq:anomaly}
 \begin{split}
  A_\sigma &= \frac{1}{2\pi}\int {\rm d}^2 x \sqrt{g}\Bigl\{
  \sigma \frac{\beta_\Phi }{2} \hat{R}-\sigma \frac{\chi_{ij}}{2}  \partial_\mu \lambda^i \partial^\mu \lambda^j
  \\
  &
  -\partial_\mu \sigma w_i \partial^\mu \lambda^i
  +\sigma \beta_\Psi \nabla_\mu S^\mu 
  +\sigma \frac{\beta^S_2}{2} S_\mu S^\mu
  \\
  &
  - \partial_\mu \sigma \beta^S_3 S^\mu
  + \sigma z_i \partial_\mu \lambda^i S^\mu 
 \Bigr\}\,.
 \end{split}
\end{equation}
The above parametrization includes all possible terms constructed with up to two derivatives of the sources.\footnote{%
Having in mind a fundamental field realization, we have also neglected possible terms that are zero on-shell, in fact there would a contribution proportional to the equations of motion off-shell \cite{Fortin:2012hn}.}
We have included the following general functions of $\lambda^i$:
$\beta_\Phi$, $\beta_\Psi$, $\beta_2^S$, $\beta^S_3$, $w_i$, $z_i$ and $\chi_{ij}$,
all of which can be interpreted as ``tensors'' in the space of couplings (scalars, vectors and a rank-$2$ symmetric tensor, respectively).\footnote{%
Whether these quantities are scalars, vectors and tensors is here decided by their Latin indices as in Refs.~\cite{Osborn:1991gm,Jack:2013sha}. This statement could be made more precise adopting a framework
in which the anomaly is manifestly invariant under couplings reparametrizations $\lambda'=\lambda'(\lambda)$, such as the one of Ref.~\cite{Andriolo:2022lcb}, which requires a further modification of the covariant derivative generated by the pull-back of the Levi-Civita connection of $\chi_{ij}$.
}
In the first line of \eqref{eq:anomaly} there are all the ``standard'' local rg contributions (for a nongauged symmetry)
and in the second line there are all the new contributions due to the presence of the source $S_\mu$. The scalars $\beta_\Phi$, $\beta_\Psi$ and $\beta^S_m$ (for $m=2,3$) are scalar functions of $\lambda^i$, the vectors $w_i$ and $z_i$ also depend on $\lambda^i$, and $\chi_{ij}$ could be interpreted as a metric in the space of couplings \cite{Jack:2013sha}. We colloquially refer to the scalar functions as ``charges'' in the following.

The computation of the Wess-Zumino consistency condition \eqref{eq:wz-consistency-general}, with $\Delta_\sigma=\Delta^W_\sigma-\Delta^\beta_\sigma$ defined in \eqref{eq:full-delta},
requires only the infinitesimal gauged Weyl transformations 
\begin{equation}
 \begin{split}
 &\delta_\sigma g_{\mu\nu} = 2 \sigma g_{\mu\nu}\,,
 \qquad
 \delta_\sigma R =-2\sigma R -2 \nabla^\mu \partial_\mu \sigma\,,
 \\&
 \delta_\sigma \nabla^\mu S_\mu = -\sigma \nabla^\mu S_\mu - \nabla^\mu \partial_\mu \sigma\,,
 \end{split}
\end{equation}
from which it is easy to see that $\delta_\sigma \hat{R} = -2\sigma \hat{R}$,
so $\sqrt{g}\hat{R}$ is invariant.
Integrating some derivatives by part, we find
\begin{equation}
 \begin{split}
  [\Delta_\sigma,\Delta_{\sigma'}] \Gamma
  = \frac{1}{2\pi}\int {\rm d}^2x \sqrt{g}\bigl(\sigma \partial_\mu \sigma'-\sigma'\partial_\mu \sigma\bigr) {\cal Z}^\mu =0\,.
 \end{split}
\end{equation}
The vector ${\cal Z}^\mu$ is a relatively long function of all the involved currents and the tensors of \eqref{eq:anomaly} that we give in a moment after having discussed its structure.

Using the chain rule $\partial_\mu = \partial_\mu \lambda^i \partial_i$ on all the tensors in the space of the couplings, it is easy to argue that ${\cal Z}_\mu$ must have the form
\begin{equation}
 \begin{split}
  {\cal Z}_\mu = \partial_\mu \lambda^i {\cal Y}_i + S_\mu {\cal X}\,,
 \end{split}
\end{equation}
so, for arbitrary $\partial_\mu \lambda^i$ and $S_\mu$, we have that ${\cal Z}_\mu=0$ implies ${\cal Y}_i=0$ and ${\cal X}=0$ independently.
The explicit forms of the two contributions to ${\cal Z}_\mu$ are
\begin{equation}\label{eq:consistency-conditions}
 \begin{split}
  {\cal Y}_i &=
  -\partial_i \beta_\Psi
  +\chi_{ij}\beta^j
  - \beta^j\partial_j w_i
  - w^j \partial_i \beta_j + z_i \,,
  \\
  {\cal X} &= \beta^S_2 -\beta^i\partial_i \beta^S_3 - z_i \beta^i \,,
 \end{split}
\end{equation}
and they are the first main result of this section.

In order to understand the implications of the consistency conditions, we define a new charge
\begin{equation}\label{eq:tilde-beta-psi-def}
 \begin{split}
  {\tilde{\beta}}_\Psi = \beta_\Psi + w_i \beta^i +\beta^S_3\,.
 \end{split}
\end{equation}
In the definition, we see that the charge $\beta_\Psi$ is shifted by a term proportional to the beta functions and the vector $w_i$ in a way familiar to the traditional local rg \cite{Osborn:1991gm},
but also by the term $\beta^S_3$ which is included for future convenience.
From ${\cal Y}_i=0$, we see that the gradient
\begin{equation}
 \begin{split}
  \partial_i {\tilde{\beta}}_\Psi =  \chi_{ij}\beta^j + (\partial_i w_j-\partial_j w_i) \beta^j
  + \partial_i \beta^S_3
  +z_i
 \end{split}
\end{equation}
has symmetric and antisymmetric terms contracting $\beta^i$ like in the standard local rg, besides two additional contributions.
The flow of ${\tilde{\beta}}_\Psi$
is almost gradient-like and the local scale derivative
$\mu\frac{{\rm d}}{{\rm d}\mu}{\tilde{\beta}}_\Psi = \beta^i\partial_i {\tilde{\beta}}_\Psi$ becomes
\begin{equation}
 \begin{split}
  \mu\frac{{\rm d}}{{\rm d}\mu} {\tilde{\beta}}_\Psi
  =\beta^i\partial_i {\tilde{\beta}}_\Psi
  =\chi_{ij}\beta^i\beta^j
  +\beta^i\partial_i \beta^S_3
  +\beta^iz_i
  \,.
 \end{split}
\end{equation}
Crucially, we can now use the solution of ${\cal X}=0$, that is, $\beta^S_2=\beta^i\partial_i \beta^S_3 + z_i \beta^i$, to simplify further
\begin{equation}\label{eq:flow-of-betapsi}
 \begin{split}
  \mu\frac{{\rm d}}{{\rm d}\mu}{\tilde{\beta}}_\Psi
  =
  \chi_{ij}\beta^i\beta^j +\beta^S_2
  \,.
 \end{split}
\end{equation}
We deduce that the function $\tilde{\beta}_\Psi$ is monotonic if $\chi_{ij}$ is positive definite and, importantly, if $\beta^S_2>0$.
For a proof of monotonicity of the rg, it would be sufficient to find at least one scheme for which $\tilde{\beta}_\Psi$ is monotonic. For example a scheme in which $\chi_{ij}>0$ and $\beta^S_2=0$ gives a corresponding $\tilde{\beta}_\Psi$ which is a candidate $C$-function \cite{Zamolodchikov:1986gt}.
We thus return on the analysis of \eqref{eq:flow-of-betapsi} after having addressed how the quantities actually transform under a change of renormalization scheme,
to see if it is possible to find a general scheme in which $\beta^S_2=0$,
but we anticipate now that it is actually not possible.

\subsection{Scheme change transformations}\label{sect:scheme}

A change in the renormalization prescription can be encoded in the redefinition of the renormalized couplings of the effective action $\Gamma$ of the local theory. This corresponds to $\Gamma \to \Gamma +\delta\Gamma$,
that we parametrize as
\begin{equation}\label{eq:scheme-change}
 \begin{split}
  \delta\Gamma = \frac{1}{2\pi}\int {\rm d}^2x \sqrt{g}
  \Bigl\{&
  \frac{b_\Phi}{2}\hat{R}
  -\frac{c_{ij}}{2}\partial_\mu \lambda^i \partial^\mu g^j
  +b_\Psi \nabla^\mu S_\mu
  \\&
  + \frac{b^S_2}{2} S^\mu S_\mu + e_i \partial_\mu \lambda^i S^\mu
  \Bigr\}
 \end{split}
\end{equation}
and all the newly introduced tensors are functions of $\lambda^i$.
From $\Gamma \to \Gamma +\delta\Gamma$, we have a corresponding change in the anomaly $A_\sigma \to A_\sigma+\delta A_\sigma$, where $\delta A_\sigma =(\Delta^W_\sigma-\Delta^\beta_\sigma)\delta\Gamma$.
Comparing with the structure of the original anomaly \eqref{eq:anomaly}, we can read how the anomaly coefficients change with the scheme.
Some tensors transform as Lie derivatives in the space of couplings
\begin{equation}\label{eq:scheme-change-1}
 \begin{split}
  \beta_\Phi
  &\to \beta_\Phi + \beta^i \partial_i b_\Phi = \beta_\Phi + {\cal L}_\beta b_\Phi
  \\
  \beta_\Psi
  &\to
  \beta_\Psi+{\cal L}_\beta b_\Psi
  \\
  \chi_{ij}
  &\to
  \chi_{ij}+\beta^l\partial_l c_{ij} + \partial_i \beta^l c_{lj} + \partial_j \beta^l c_{il}
  =\chi_{ij}+{\cal L}_\beta c_{ij}
  \\
  \beta^S_2
  &\to \beta^S_2 +{\cal L}_\beta b^S_2
  \\
  z_i
  &\to
  z_i + \beta^j\partial_j e_i +  e_j \partial_i \beta^j
  = z_i +{\cal L}_\beta e_i \,,
 \end{split}
\end{equation}
where ${\cal L}_\beta$ is the Lie derivative in the ``direction'' of the vector $\beta^i$ acting on scalars, a symmetric tensor and a form. The
others transform with ``shifts''
\begin{equation}\label{eq:scheme-change-2}
 \begin{split}
  w_i
  &\to w_i +e_i+c_{ij}\beta^j -\partial_i b_\Psi
  \\
  \beta^S_3
  &\to \beta^S_3 +b^S_2 -e_i \beta^i \,.
 \end{split}
\end{equation}
It is easy to check that the consistency equations, ${\cal Y}_i=0$ and ${\cal X}=0$, are invariant under the combination of the transformations \eqref{eq:scheme-change-1} and \eqref{eq:scheme-change-2}.
As for the quantity of interest, using the definition $\tilde{\beta}_\Psi = \beta_\Psi +w_i\beta^i +\beta^S_3$, we have
\begin{equation}\label{eq:tilde-beta-psi-transf}
 \begin{split}
  \tilde{\beta}_\Psi
  & \to \tilde{\beta}_\Psi 
  + c_{ij} \beta^i \beta^j + b^S_2\,,
 \end{split}
\end{equation}
that transforms similarly to the local rg counterpart \cite{Osborn:1991gm}, except for the additional contribution $b^S_2$.

The transformations induced by the choice of scheme cannot be used to set the charge $\beta^S_2$ to zero, because $\beta^S_2 \to \beta^S_2 +{\cal L}_\beta b^S_2$, unless $\beta^S_2$ has the form of the gradient of a function contracted with the beta functions.
In other words, we \emph{cannot} find a general scheme for an arbitrary model in which $\beta^S_2$ is zero and this fact will be confirmed by the explicit examples of Sect.~\ref{sect:examples}.
Nevertheless, we discuss the self-consistency of the special case $\beta^S_2=0$
along the rg flow in the next subsection.

\subsection{Setting the charge $\beta^S_2=0$ consistently}\label{sect:setting}

We can investigate the interplay of the special case $\beta^S_2=0$ along the flow with the Wess-Zumino consistency conditions. If we \emph{impose} $\beta^S_2=0$,
we have from Eq.~\eqref{eq:consistency-conditions} that $\beta^i\partial_i \beta^S_3 = - z_i \beta^i$, which implies, for general $\beta^i$, that
\begin{equation}
 \begin{split}
  z_i=-\partial_i \beta^S_3 -N_i\,,
 \end{split}
\end{equation}
where $N_i$ is a form in the cotanget space to the couplings orthogonal to $\beta^i$, i.e., $N_i\beta^i=0$.
Using this relation, we see that the last two terms of the anomaly \eqref{eq:anomaly} become
\begin{equation}\label{eq:anomaly-red}
 \begin{split}
  A_\sigma &\supset \frac{1}{2\pi}\int {\rm d}^2 x \sqrt{g}\Bigl\{
  - \partial_\mu \sigma \beta^S_3 S^\mu
  - \sigma \partial_i \beta^S_3 \partial_\mu \lambda^i S^\mu
  \\&\qquad\qquad\qquad\qquad
  - \sigma N_i \partial_\mu \lambda^i S^\mu
 \Bigr\}
 \\
 &=
 -\frac{1}{2\pi}\int {\rm d}^2 x \sqrt{g}\Bigl\{
  \partial_\mu (\sigma \beta^S_3) S^\mu
  +\sigma N_i \partial_\mu \lambda^i S^\mu
 \Bigr\}
 \,.
 \end{split}
\end{equation}
Integrating by parts the first term, we thus have that in the case $\beta^S_2=0$
the complete anomaly becomes
\begin{equation}\label{eq:anomaly-simplified}
 \begin{split}
  A_\sigma &=
 \frac{1}{2\pi}\int {\rm d}^2 x \sqrt{g}\Bigl\{
  \sigma \frac{\beta_\Phi }{2} \hat{R}-\sigma \frac{\chi_{ij}}{2}  \partial_\mu \lambda^i \partial^\mu g^j
  \\&
  +\sigma \Bigl(\beta_\Psi +\beta^S_3 \Bigr) \nabla_\mu S^\mu
  -\partial_\mu \sigma w_i \partial^\mu \lambda^i
  -\sigma N_i \partial_\mu \lambda^i S^\mu
 \Bigr\}
 \,.
 \end{split}
\end{equation}
The anomaly term $\nabla^\mu S_\mu$ has the additional additive contribution $\beta^S_3$, which reproduces the term included in the previous definition of $\tilde{\beta}_\Psi$ given in \eqref{eq:tilde-beta-psi-def}. In fact $\beta^S_3$ is no longer necessary and we could define $\beta'_\Psi = \beta_\Psi+\beta^S_3$, so that the anomaly becomes
\begin{equation}\label{eq:anomaly-simplified2}
 \begin{split}
  A_\sigma &= 
 \frac{1}{2\pi}\int {\rm d}^2 x \sqrt{g}\Bigl\{
  \sigma \frac{\beta_\Phi }{2} \hat{R}-\sigma \frac{\chi_{ij}}{2}  \partial_\mu \lambda^i \partial^\mu g^j
  \\&\qquad\quad
  +\sigma \beta'_\Psi \nabla_\mu S^\mu
  -\partial_\mu \sigma w_i \partial^\mu \lambda^i
  -\sigma N_i \partial_\mu \lambda^i S^\mu
 \Bigr\}
 \,.
 \end{split}
\end{equation}
Following the analysis of Sect.~\ref{sect:two} we can see that the charge
$\tilde{\beta}_\Psi = \beta'_\Psi + w_i \beta^i $ is now such that
\begin{equation}\label{eq:flow-of-betapsi-simplified}
 \begin{split}
  \mu\frac{{\rm d}}{{\rm d}\mu}{\tilde{\beta}}_\Psi
  =
  \chi_{ij}\beta^i\beta^j
  \,,
 \end{split}
\end{equation}
and has the same structure of the rg flow given for the charge $\tilde{\beta}_\Phi$ by Osborn in Ref.~\cite{Osborn:1991gm}.
At this stage, for monotonicity one wound only need to find a scheme in which $\chi_{ij}>0$, which is possible only if the underlying theory is unitary.

The choice $\beta^S_2=0$ must be consistent with the choice of scheme, so the available changes of scheme given in \eqref{eq:scheme-change}
must also be such that $b^S_2=0$, implying that we always remain in a scheme without $S_\mu S^\mu$ anomaly. As a consequence, if we remain in such schemes with $b^S_2=0$, we have that the transformation \eqref{eq:tilde-beta-psi-transf} becomes
\begin{equation}\label{eq:tilde-beta-psi-transf-simplified}
 \begin{split}
  \tilde{\beta}_\Psi
  & \to \tilde{\beta}_\Psi 
  + c_{ij} \beta^i \beta^j \,,
 \end{split}
\end{equation}
which also agrees with the transformation of $\tilde{\beta}_\Phi$ given in Ref.~\cite{Osborn:1991gm}.
Using \eqref{eq:scheme-change-1} and \eqref{eq:scheme-change-2}, we see that choice $\beta^S_2=0$ also implies
\begin{equation}\label{eq:scheme-change-3}
 \begin{split}
  z_i
  &\to
  z_i + \beta^j\partial_j e_i +  e_j \partial_i \beta^j
  = z_i +{\cal L}_\beta e_i \,,
  \\
  \beta^S_3
  &\to \beta^S_3 -e_i \beta^i \,.
 \end{split}
\end{equation}
Recall now that the case $\beta^S_2=0$ implies that $z_i =-\partial_i\beta^S_3-N_i$, so the two transformations above cannot actually be independent. In fact, from their combination, we can deduce a transformation for the $N_i$. From the transformation of $\beta^S_3$ we infer $\partial_i\beta^S_3 \to \partial_i\beta^S_3 -\partial_i(e_j \beta^j) = \partial_i\beta^S_3 -\beta^j \partial_i e_j -e_j \partial_i \beta^j$. We thus have
\begin{equation}\label{eq:scheme-change-4}
 \begin{split}
  N_i=-z_i-\partial_i\beta^S_3
  &\to
  N_i +\beta^j \partial_i e_j 
  -\beta^j\partial_j e_i \,.
 \end{split}
\end{equation}
Geometrically, $N=N_i{\rm d}\lambda^i$ is a one-form in the space of coupling and the transformation on the rhs of \eqref{eq:scheme-change-4}
depends on the exterior differential ${\rm d}e$ of $e=e_i {\rm d}\lambda^i$, which is a two-form. In this notation, we can write
\begin{equation}
 \begin{split}
  N
  &\to
  N+ {\rm d} e (\beta,\cdot)
 \end{split}
\end{equation}
where $\beta=\beta^i\partial_i$ is a vector in the space of couplings and
the two-form is evaluated on only one argument. By construction $N(\beta)=N_i\beta^i$ remains zero along a scheme transformation because ${\rm d} e (\beta,\beta)=0$, as expected. Using Cartan's formula
\begin{equation}
 \begin{split}
  N
  &\to
  N+ {\cal L}_\beta e - {\rm d} (e_i \beta^i)\,.
 \end{split}
\end{equation}
This is telling us that the part of $e_i$ that is orthogonal to $\beta^i$ transforms the vectors $N_i$ homogeneously, like the transformations \eqref{eq:scheme-change-1},
while the rest transforms the $n_i$ inhomogeneously, like the transformations \eqref{eq:scheme-change-2}.

\subsection{Comparison with the standard local rg}\label{sect:comparison}

As a consequence of the analysis of Sects.~\ref{sect:scheme} and, in part, of Sect.~\ref{sect:setting} we now know that it is not possible to find a general scheme in which $\beta^S_2=0$. This result is somehow expected, because we know that conformal invariance and unitarity imply irreversibility of the rg flow, but there is no reason to expect that scale invariance alone implies irreversibility.
Since conformal invariance is a special case of scale invariance, our discussion points to the fact that the ``signature'' of pure scale invariance in the rg flow
is precisely the charge $\beta^S_2$. We thus expect that $\beta^S_2$ is nonzero for models that are scale invariant, but not conformal invariant, which is confirmed by the examples of the next sections.

The standard results of the local rg can be reproduced ``trivially'' by decoupling the gauge potential $S_\mu$ from the anomaly. This is achieved by requiring $\beta_\Psi = \beta_\Phi$, which cancels the divergence in $A_\sigma$, $\beta^S_{m}=0$ for $m=2,3$ and $z_i=0$
in \eqref{eq:anomaly}. In this case there must be no change in the renormalization of the $S$-independent contributions, so a different scheme can be achieved only for $b_\Psi=b_\Phi$ and for $b^S_2=e_i=0$ in \eqref{eq:scheme-change}. With this in mind, we see that $\tilde{\beta}_\Psi$
becomes $\tilde{\beta}_\Phi=\beta_\Phi + w_i \beta^i$
considered in Ref.~\cite{Osborn:1991gm}.

Then, if and only if $S_\mu$ is decoupled, one could follow the proof of Ref.~\cite{Osborn:1991gm} that $\tilde{\beta}_\Phi$ is monotonic, or, more precisely, that there exists at least one scheme for which $\chi_{ij}$
is positive definite for unitary theories.
The scheme of choice is the one in which $\chi_{ij}$
is Zamolodchikov's metric, $G_{ij}= \left|x\right|^4\langle{\cal O}_i(x){\cal O}_j(0)\rangle$, and $\tilde{\beta}_\Phi$ becomes Zamolodchikov's monotonic $C$-function up to a constant, completing the elegant proof of irreversibility of \cite{Osborn:1991gm}. In general, $\beta_\Phi$ itself differs from the $C$-function by terms proportional to $\beta^i$ modulo a normalization, so the two quantities
coincide at fixed points of the renormalization group.\footnote{%
The fact that $\chi_{ij}$ can be related to Zamolodchikov's metric $\langle{\cal O}_i {\cal O}_j \rangle$ is a luxury of the two dimensional case. In four dimensions there is a consistency condition that leads to a result similar to \eqref{eq:flow-of-betapsi},
but in that case the symmetric tensor relates to the correlator $\langle{\cal O}_i {\cal O}_j T\rangle$, for which it is not obvious to establish positivity \cite{Osborn:1991gm,Fortin:2012hn}.
}

Returning to the case of nonzero gauge potential $S_\mu$, things become more complicate. There is still the freedom of changing the scheme, but there is no general scheme in which $\beta^S_2$ is zero as seen in Sect.~\ref{sect:scheme},
so monotonicity is not guaranteed because of the extra contribution in \eqref{eq:flow-of-betapsi}, even if the underlying theory is unitary.
The simplest solution would be to find some scheme for which $\beta^S_2=0$, but this can only be done with a specific model in mind, though we discussed the consistency of this choice along the flow in Sect.~\ref{sect:setting}. In this case, from \eqref{eq:flow-of-betapsi}, we could deduce $\mu\frac{{\rm d}}{{\rm d}\mu}{\tilde{\beta}}_\Psi=\chi_{ij}\beta^i\beta^j$
and eventually develop a proof to monotonicity that could apply to special models
following the steps of Ref.~\cite{Osborn:1991gm}, assuming that such models are also unitary (i.e., such that there is a scheme in which $\chi_{ij}>0$). Notice, however, that all the famous examples of scale-but-not-conformal theories are nonunitary theories \cite{Gimenez-Grau:2023lpz}, which leaves little hope for a generalization of Zamolodchikov's theorem to even a subclass of scale invariant theories.
That said, the consistency of the choice $\beta^S_2=0$ involves normal terms to the beta functions in the anomaly, but they do not seem to have much effect in the form of the flow.

For a final remark, notice that there is no consistency condition involving $\beta_\Phi$,
because $\beta_\Phi$ is coupled to a Weyl-invariant density (see also the discussion of Appendix~\ref{sect:geometry}). This may be counterintuitive since the local rg without the gauging of the Weyl potential constrains the coefficient of the scalar curvature anomaly, but it actually happens because of our parametrization, having coupled $\beta_\Phi$ to $\hat{R}=R-2\nabla^\mu S_\mu$, which now plays the role of a type-B anomaly, instead of $R$ which is a type-A anomaly \cite{Deser:1993yx}.

\subsection{Boundary terms in two dimensions}\label{sect:boundary}

We now want to succintly add to the considerations of Sect.~\ref{sect:two}
the effects of a finite boundary in the Weyl anomaly. We are still working in two dimensions, meaning that the boundary is a one dimensional submanifold
with some coordinate $\tau$ orthogonal to a vector $n^\mu$ of unit norm, which we assume to exist in a neighbor of the boundary itself.
The intrinsic geometry of the boundary is characterized by an ``induced'' metric from projecting $g_{\mu\nu}\frac{{\rm d}x^\mu}{{\rm d}\tau}\frac{{\rm d}x^\nu}{{\rm d}\tau}$, and it is convenient to parametrize the measure on the boundary with ${\rm d}\tau={\rm d}s$, where $s$ is the arc-length measured using $g_{\mu\nu}$. There is also an extrinsic component to the geometry, given here by the trace of the extrinsic curvature tensor, $K=g^{\mu\nu}\nabla_\mu n_\nu$. Using the covariant derivative $\hat{\nabla}$, it is natural to define $\hat{K}=K +n^\mu S_\mu$, which has the meaning of extrinsic curvature of the gauged connection. Under infinitesimal Weyl transformations we have
\begin{equation}
 \begin{split}
 &\delta_\sigma n^\mu = -\sigma n^\mu\,,
 \qquad \delta_\sigma {\rm d}s =\sigma {\rm ds}\,,
 \\&
 \delta_\sigma K = -\sigma K - n^\mu \partial_\mu \sigma\,,
 \end{split}
\end{equation}
from which it is easy to see that $\delta_\sigma \hat{K} = -\sigma \hat{K}$,
so $\hat{K}$ transforms covariantly and the combination ${\rm d}s \hat{K}$ is invariant similarly to $\sqrt{g}\hat{R}$ in Sect.~\ref{sect:two}.

In presence of a finite boundary the effective action must contain a new boundary contribution $\Gamma_{\rm tot}= \Gamma+\Gamma_{\rm bd}$.
For consistency, to the anomaly \eqref{eq:anomaly} we must add a boundary term, which we parametrize as
\begin{equation}\label{eq:boundary-anomaly}
 \begin{split}
 A_{{\rm bd},\sigma}
 = \frac{1}{2\pi}\int {\rm d}s \Bigl\{&
 \sigma \hat{\beta}_\Phi \hat{K}
 +\sigma \omega_i n^\mu \partial_\mu \lambda^i
 \\
 &
 +\varepsilon n^\mu \partial_\mu \sigma
 -\sigma \hat{\beta}_\Psi n^\mu S_\mu
 \Bigr\}\,.
 \end{split}
\end{equation}
For parametrizing the boundary we have made similar choices as we have done for the bulk in \eqref{eq:anomaly}, particularly in regards to the use of $\hat{K}$ as monomial. Similarly to the bulk, we have that $\hat{\beta}_\Phi$,
$\hat{\beta}_\Psi$, $\varepsilon$ are scalar functions of $\lambda^i$, and that $\omega_i$ is a vector form in the space of the couplings. An anomaly of the form $\partial_\mu \sigma S^\mu$ is not included because it can be integrated by parts and, by construction, it does not have any additional effect given that the boundary of the boundary is empty.

New terms appear when applying the Wess-Zumino consistency condition to $A_{{\rm tot},\sigma}=A_\sigma+A_{{\rm bd},\sigma}$. The new contributions come from applying $\Delta_\sigma$ directly to the boundary terms, but also from the total derivatives in the bulk that have been neglected in Sect.~\ref{sect:two}. Structurally the Wess-Zumino condition becomes
\begin{equation}
 \begin{split}
 [\Delta_\sigma,\Delta_{\sigma'}] \Gamma_{\rm tot}
  &= \frac{1}{2\pi}\int {\rm d}^2 x \sqrt{g}\bigl(\sigma \partial_\mu \sigma'-\sigma'\partial_\mu \sigma\bigr) {\cal Z}^\mu
  \\&
  + \frac{1}{2\pi}\int {\rm d}s\, n^\mu \bigl(\sigma \partial_\mu \sigma'-\sigma'\partial_\mu \sigma\bigr) {\cal B} =0\,,
 \end{split}
\end{equation}
where ${\cal Z}^\mu$ is the same vector computed in Sect.~\ref{sect:two}, and
\begin{equation}
 \begin{split}
 {\cal B} = \beta_\Psi -\hat{\beta}_\Psi  - \omega_i\beta^i+\beta^i \partial_i \varepsilon\,.
 \end{split}
\end{equation}
Consequently, ${\cal B}=0$ is the new consistency condition that is intrinsic to the boundary.
As happened for $\beta_\Phi$ in Sect.~\ref{sect:two}, there is
no condition on $\hat{\beta}_\Phi$, because it couples to a type-B anomaly.
The interesting point is that the charges $\beta_\Psi$ and $\hat{\beta}_\Psi$ are equal at fixed points, i.e., $\hat{\beta}_\Psi=\beta_\Psi$ from ${\cal B}=0$ for $\beta^i=0$. We can also recover the original local rg result of Ref.~\cite{Osborn:1991gm} when decoupling $S_\mu$, which, in this case, corresponds to the requirement $\hat{\beta}_\Psi=\hat{\beta}_\Phi$
and the relation ${\cal B}=0$ then reproduces the boundary consistency condition given in 
Ref.~\cite{Osborn:1991gm}.

The boundary terms of the anomaly can be parametrized in a new scheme by changing $\Gamma_{{\rm bd}} \to \Gamma_{{\rm bd}}+\delta \Gamma_{{\rm bd}}$, where
\begin{equation}
 \begin{split}
 \delta \Gamma_{\rm bd} =
 \frac{1}{2\pi}\int {\rm d}s \Bigr\{
 \hat{b}_\Phi \hat{K}
 +n^\mu d_i \partial_\mu \lambda^i
 -\hat{b}_\Psi n^\mu S_\mu
 \Bigr\}\,,
 \end{split}
\end{equation}
and, similarly to Sect.~\ref{sect:boundary}, we have chosen conventions that simplify the final form of the transformations below.
To obtain how the tensors of \eqref{eq:boundary-anomaly} transform, we must use the definition, $A_\sigma = \Delta_\sigma \Gamma$, which induces the transformation 
$A_{{\rm bd},\sigma} \to A_{{\rm bd},\sigma}+\delta A_{{\rm bd},\sigma}$.
Notice also that the transformation of the bulk does not affect the boundary
in the general case. Finally, the tensors of \eqref{eq:boundary-anomaly} transform as
\begin{align}
 & \hat{\beta}_\Phi \to \hat{\beta}_\Phi + {\cal L}_\beta \hat{b}_\Phi
 && \hat{\beta}_\Psi \to \hat{\beta}_\Psi + {\cal L}_\beta \hat{b}_\Psi
 \nonumber\\
 &\omega_i \to \omega_i + {\cal L}_\beta d_i
 && \varepsilon \to \varepsilon - \hat{b}_\Psi +d_i \beta^i\,,
\end{align}
and it is easy to see that ${\cal B}$ is invariant under the above transformations. Arguably, we could use the freedom of the choice of the scheme to parametrize the anomaly so that $\varepsilon=0$. The analysis of this section could be relevant to open string theory, should the need of the gauged version of Weyl symmetry arise.

\section{Examples}\label{sect:examples}

We want to consider nontrivial yet simple examples to tie things together.
We choose two that are motivated in part by the analysis of the charge $\beta^S_2$,
since it is the main actor in differentiating \eqref{eq:flow-of-betapsi}
from the standard local rg result of Ref.~\cite{Osborn:1991gm}.
The first example is a higher derivative free scalar theory that does not admit a standard Weyl covariant description in curved space because of a geometrical obstruction, yet it is a CFT in flat space \cite{Brust:2016gjy}.
The second one is the theory of elasticity, which is scale invariant but not conformal invariant in flat space as discussed by Riva and Cardy \cite{Riva:2005gd}.
The two examples display zero and nonzero values for $\beta^S_2$, which, we argue, occur in relation to their conformal properties.

\subsection{The free higher-derivative scalar and $\beta^S_2=0$}\label{sect:example}

To begin with we restrict our attention to ``generalized free''
fields with quadratic actions coupled to a Weylian geometry, playing the role of ``Gaussian'' fixed points. 
A standard scalar with a two derivatives action is not particularly interesting for us, because the scalar has canonical dimension $(d-2)/2$, so its Weyl weight is zero in $d=2$ and it does not couple minimally with $S_\mu$. A standard Dirac spinor does not work either, because even if it does couple to $S_\mu$ through $\hat{\slashed{\nabla}}$, we have that, due to the (anti)hermiticity of the Dirac Lagrangian, $S_\mu$ cancels in a symmetrized action.\footnote{%
The simple curved space action with Lagrangian ${\cal L} \sim -\varphi \nabla^2 \varphi$ is already conformally invariant in $d=2$, because $-\nabla^2$ is conformally covariant in $d=2$. 
Furthermore, the presence of $S_\mu$ in the Dirac operator can be interpreted as a contribution of the vector part of the torsion up to a factor $1/(d-1)$ \cite{Karananas:2015eha,Sauro:2022chz}, which decouples from a Dirac spinor. In other words, also $\slashed{\nabla}$ is a conformally covariant operator and this is true in any $d$.
For a complete discussion of the degrees of freedom of the torsion tensor see Ref.~\cite{Martini:2023rnv}.
}

The above considerations motivate the simplest nontrivial example that works for our purpose, i.e., a \emph{higher} derivative scalar $\varphi$, with canonical dimension $\frac{d-4}{2} \to -1$ and thus Weyl weight $w_\varphi=\frac{4-d}{2} \to 1$. We first concentrate on the computation of the anomaly and then discuss its relevance more pragmatically. The Weyl gauged classical action
\begin{equation}\label{eq:4der-action}
 \begin{split}
  S[\varphi,g_{\mu\nu},S_\mu]=-\frac{1}{2}\int {\rm d}^dx \sqrt{g} \varphi (\hat{\nabla}^2)^2 \varphi\,,
 \end{split}
\end{equation}
is manifestly invariant under \eqref{eq:gauged-weyl-transformations}
and $\varphi \to {\rm e}^{w_\varphi \sigma }\varphi$ in any $d$, including $d=2$, given the covariant derivative $\hat{\nabla}$ that is defined in \eqref{eq:nabla-hat}. We set on $d=2$ for the explicit computation below, but notice that the field has then \emph{negative} scaling dimension, signalling that it is related to a special type of CFT \cite{Brust:2016gjy}.

Manifest gauge invariance is lost if the covariant derivative is expressed in terms of the Levi-Civita connection. We have that the ``kinetic'' operator is of the form
\begin{equation}\label{eq:hd-operator}
 \begin{split}
  (\hat{\nabla}^2)^2 \varphi
  = (\nabla^2)^2 \varphi
  +B^{\mu\nu} \nabla_\mu \partial_\nu \varphi
  +C^\mu \partial_\nu \varphi
  +D \varphi
  \,,
 \end{split}
\end{equation}
with relatively simple tensor coefficients in $d=2$
\begin{equation}
 \begin{split}
 B_{\mu\nu} &=
 2 g_{\mu\nu} S^\rho S_\rho -4 S_\mu S_\nu + 4 \nabla_{(\mu} S_{\nu)}\,,
 \\
 C_\mu
 &=
 2 R_{\mu\nu} S^\nu-4 S_\mu \nabla^\nu S_\nu
 +4 S^\nu W_{\mu\nu}
 \\&\quad
 +2 \nabla^2 S_\mu
 +2 \nabla_\mu \nabla^\nu S_\nu \,,
 \end{split}
\end{equation}
and $D=(S^\mu S_\mu)^2 + \cdots$, where the dots hide several terms involving $S_\mu$ and its derivatives that we do not report for indolence. Crucially, there is no term with three derivatives.

Classically the energy-momentum tensor and the dilation current are complicate functions of the field $\varphi$ and the sources. We have computed them in general, but, for brevity, we give them off-shell in the flat space limit
\begin{equation}
 \begin{split}
 T^{\mu\nu} &= 2 \partial^2\partial^{(\mu}\varphi\partial^{\nu)}\varphi
 -\delta^{\mu\nu} \partial^2\partial^{\lambda}\varphi\partial_\lambda\varphi
 -\frac{1}{2} \delta^{\mu\nu}(\partial^2\varphi)^2\,,
 \\
 D^\mu &= -\partial^2\varphi \partial^\mu\varphi+\varphi \partial^2 \partial^\mu\varphi\,,
 \end{split}
\end{equation}
and it is straightforward to check that $T=\partial_\mu D^\mu$ using the flat space limit of the equations of motion $(\partial^2)^2\varphi=0$.
Off-shell, the relation becomes $T-\partial_\mu D^\mu=\varphi (\partial^2)^2 \varphi$, where on the rhs we have the flat space limit of $\frac{\delta}{\delta\sigma}\int w_\varphi \sigma \varphi \frac{\delta S}{\delta\varphi}$,
in agreement with N\"other theorem.
Notice that it is the contribution of the dilation current, and not the one of the trace of the energy-momentum tensor, that brings the
term proportional to the equations of motion of $\varphi$ on the rhs.
It is also easy to deduce that $D^\mu$ is not the gradient of a scalar, nor the
divergence of a symmetric tensor on-shell, because such symmetric tensor would have two derivatives and the equations of motion cannot help in its construction. This is consistent with the fact that the energy-momentum tensor does not admit an improvement to tracelessness and is related to a geometrical obstruction that we mention below.

Using heat kernel methods \cite{Vassilevich:2003xt}, we can easily compute the anomaly without having to introduce further external sources, e.g., we compute what is occasionally referred to as the vacuum functional. More details of the procedure, which is based on heat kernel methods, can be found in Appendix~\ref{sect:hk}.
The computation is greatly simplified by the fact that we work in $d=2$, in fact we can use the results of Ref.~\cite[Sect.~VI]{Barvinsky:2021ijq} to show that the anomaly depends only on the trace of the tensor $B_{\mu\nu}$ and on the curvature scalar $R$
\begin{equation}\label{eq:hk-computation}
 \begin{split}
 A_\sigma & = \frac{1}{2\pi}\int {\rm d}^2 x \sqrt{g} \sigma \Bigl\{
 \frac{R}{6}+\frac{1}{4} B^\mu{}_\mu
 \Bigr\}
 \\
 &
 = \frac{1}{2\pi}\int {\rm d}^2 x \sqrt{g} \sigma \Bigl\{
 \frac{R}{6}+\nabla^\mu S_\mu
 \Bigr\}\,.
 \end{split}
\end{equation}
By comparison with \eqref{eq:anomaly}, we deduce the nonzero values of the charges $\beta_\Phi=\frac{1}{3}$ and $\beta_\Psi=\frac{4}{3}$, which should be understood as ``Gaussian'' fixed point values of some rg flow.
There is no $S^2$ anomaly, i.e., $\beta^S_2=0$, so it is consistent to discuss schemes without such charge for this model as in Sect.~\ref{sect:setting}. Furthermore,
we have that $\beta_\Phi=\frac{1}{3}$ is twice as much the contribution of a simple scalar (which would be $\beta_\Phi=\frac{1}{6}$), and this is what one could expect from a simple higher derivative scalar and a one-loop analysis.

A practical reason why the action \eqref{eq:4der-action} is interesting is because
in even dimensions there are obstructions to extending some flat space's CFTs to fully Weyl and diffeomorphisms invariant models.
The simplest example of obstruction is seen by considering the conformally invariant four-derivatives scalar in general $d$. In curved space, its action is governed by a Weyl-covariant operator \cite{Fradkin:1982xc,Riegert:1984kt}, sometimes known as Paneitz operator in arbitrary $d$ \cite{Paneitz}. The Paneiz operator depends on the Schouten tensor
\begin{equation}\label{eq:schouten-def}
 \begin{split}
  P_{\mu\nu}= \frac{1}{d-2}\Bigl\{R_{\mu\nu}-\frac{1}{2(d-1)}R g_{\mu\nu}\Bigr\}\,,
 \end{split}
\end{equation}
which is required to balance the Weyl transformations of terms with two or more covariant derivatives of the field \cite{Paci:2023twc}. However, the tensor has a pole in $d=2$, signalling the presence of a geometrical obstruction to Weyl invariance.
Amusingly, in this Section's example we obtain $\beta^S_2=0$, which is what we would expect from any ordinary CFT and this is probably related to the fact that in flat space the model is known to be a generalized free CFT \cite{Brust:2016gjy}.

\subsection{Gauging the theory of elasticity and $\beta^S_2\neq 0$}\label{sect:cardy-riva}

Now we want to address an example of scale-but-not-conformal theory. The two best known examples are the Aharony-Fisher model of dipolar ferromagnets \cite{AF}
and the theory of elasticity, as originally pointed out by Riva and Cardy \cite{Riva:2005gd}. For both theories, the virial current cannot be expressed as a total divergence, so the models have the virial current as a special vector operator of dimension $d-1$ in their spectra \cite{Gimenez-Grau:2023lpz}.

We concentrate on the model discussed by Riva and Cardy. The action in flat space is
\begin{equation}\label{eq:cardy-riva-flat}
 \begin{split}
 S[u] & = \frac{1}{2}\int {\rm d}^2 x\Bigl\{ 2 l_1 u_{\mu\nu} u^{\mu\nu} + l_2 u_{\mu}{}^\mu u_\nu{}^\nu
 \Bigr\}\,,
 \end{split}
\end{equation}
where $u_{\mu\nu} = \partial_{(\mu}u_{\nu)}$ and $u_\mu$ is a vector describing the local deformation of some material. The coefficients $l_1$ and $l_2$ are related to shear and bulk stresses of the underlying material.\footnote{%
The coefficient $l_1$ and $l_2$ are generally denoted $\mu$ and $\lambda$, respectively, however here we do not want them confused with scale and couplings in this paper.
The action $S[u]$ is quadratic, but an interacting version describes the crystalline/tethered membrane model, which is also an example of scale-but-not-conformal model \cite{Mauri:2021ili}. Interestingly, all these models can be understood as coming from cosets that involve spacetime symmetries in the breaking process \cite{Zanusso:2014wsa}. The fact that the field variables share the indices of spacetime is what allows for more than one independent contraction in \eqref{eq:cardy-riva-flat} and it would be interesting to investigate this property in relation to the existence of a virial current.
}
The model is a scale invariant theory that, for general $l_i$ is not conformally invariant. In fact, there is no way to write down an improved (traceless and conserved) energy-momentum tensor because the virial current cannot be written as the divergence of a scalar field \cite{Riva:2005gd}. It is also not unitary.

However, in the special case $3l_1+l_2=0$, which we call ``global conformal limit'' (in the sense of Ref.~\cite{Nakayama:2016dby}),
a finite subgroup of the infinite dimensional conformal symmetry (the Virasoro group) survives. The implication of this is seen below.

Using the connection $\hat{\nabla}$ defined in Sect.~\ref{sect:classical}, it becomes trivial to write an action that generalizes \eqref{eq:cardy-riva-flat} to curved space and that is invariant under gauged Weyl transformations. The ``gauged'' elasticity model is
\begin{equation}\label{eq:cary-riva-gauged}
 \begin{split}
 S[u,g_{\mu\nu},S_\mu] & = \frac{1}{2}\int {\rm d}^2 x \sqrt{g} \Bigl\{ 2 l_1 \hat{u}_{\mu\nu} \hat{u}^{\mu\nu} + l_2 \hat{u}_{\mu}{}^\mu \hat{u}_\nu{}^\nu
 \Bigr\}
 \end{split}
\end{equation}
where now $\hat{u}_{\mu\nu} = \hat{\nabla}_{(\mu}u_{\nu)}$ with $\hat{\nabla}$ defined in \eqref{eq:nabla-hat}. We assign the Weyl weight $w_u= w(u^\mu)=-1$, which is also the natural Weyl weight that one would expect for a coordinate displacement
with units of length
(besides being the necessary choice for Weyl invariance).

According to the definitions \eqref{eq:expectation-values}, we can find a symmetric energy-momentum tensor and a dilation current satisfying $T=\nabla^\mu D_\mu$ when going on-shell using the equations of motion of the field $u^\mu$. For simplicity we give them in the flat space limit, i.e., for $g_{\mu\nu} \to \delta_{\mu\nu}$ and $S_\mu\to 0$, 
\begin{equation}
 \begin{split}
 T_{\mu\nu} = &
 4 l_1 \partial_\rho u_\mu \partial^\rho u_\nu -8 l_1 \partial_\rho u^\rho u_{\mu\nu}
 -4 l_1 \partial_\mu u^\rho \partial_\nu u_\rho
 \\&
 +4 l_1 \delta_{\mu\nu} u_{\rho\sigma}u^{\rho\sigma}
 -8l_1 u^\rho\partial_\rho u_{\mu\nu}
 -4 l_2 \delta_{\mu\nu} u^\rho\partial_\rho u^\sigma{}_\sigma
 \\&
 -2 l_2 \delta_{\mu\nu} u^\rho{}_\rho u^\sigma{}_\sigma
 \,,
 \\
 D_\mu =&
 -4(2l_1+l_2) u_\mu u^\rho{}_\rho
 +8 l_1 u^\nu u_{\mu\nu}\,,
 \end{split}
\end{equation}
but they can be computed with minimal effort in the general case.
It is easy to check that $T=\nabla^\mu D_\mu$ in general on-shell.

In the global conformal limit, $3 l_1+l_2=0$, the virial current is the divergence of a symmetric tensor
\begin{equation}
 \begin{split}
 D_\mu = \partial^\nu X_{\mu\nu}\,, \qquad X_{\mu\nu} = 4 l_1 u_\mu u_\nu + 2 l_1 \delta_{\mu\nu}u_\rho u^\rho\,,
 \end{split}
\end{equation}
having identified the virial current with the dilation current in the flat-space limit.
In $d>2$ this would imply that scale invariance could be extended to conformal invariance by appropriately improving the variational energy-momentum tensor
introducing a coupling between the symmetric tensor and the Schouten tensor (defined in \eqref{eq:schouten-def}) of the form $\int \sqrt{g} X_{\mu\nu} P^{\mu\nu}$ in the original action among other terms \cite{OsbornLectures}. However, the improvement of an energy-momentum tensor based on $X_{\mu\nu}$ truly works only in $d > 2$, because of the same obstruction discussed in the example of Sect.~\ref{sect:example}. In fact, in $d=2$ one needs the stronger condition that $D_\mu = \partial_\mu X$, where $X$ is some scalar operator. In other words, $X_{\mu\nu}= \delta_{\mu\nu} X$ must only have a trace part on-shell \cite{Polchinski:1987dy}.

We can compute the anomaly \eqref{eq:anomaly} of the gauged action of elasticity.
The simplest way to approach the computation is to use the curved space generalization of a trick applied by Nakayama in a similar context \cite{Nakayama:2016dby}, by which we define
\begin{equation}\label{eq:decomposition}
 \begin{split}
 u_\mu = \nabla_\mu \phi + \epsilon_{\mu\nu} \nabla^\nu \chi\,,
 \end{split}
\end{equation}
so we decompose the vector $u_\mu$ in a scalar and a pseudo-scalar using the covariant $\epsilon$-tensor. The decomposition assumes implicitly that there is no global zero mode (a harmonic form) and neglects the constant shift of $u_\mu$, which would not contribute to the energy balance of the statistical system anyway.

The use of \eqref{eq:decomposition} and integration by parts brings the action \eqref{eq:cary-riva-gauged} in a form that is more useful for the computation of the anomaly
\begin{equation}
 \begin{split}
 S[u,g_{\mu\nu},S_\mu] &=
 \frac{1}{2}\int {\rm d}^2 x \sqrt{g} \begin{pmatrix}
 \phi & \chi 
\end{pmatrix}
 \begin{pmatrix}
 \hat{\Delta}_{\phi\phi} & \hat{\Delta}_{\phi\chi} \\
 \hat{\Delta}_{\chi\phi} & \hat{\Delta}_{\chi\chi} 
\end{pmatrix}
\begin{pmatrix}
 \phi \\
 \chi 
\end{pmatrix}\,,
 \end{split}
\end{equation}
where $\hat{\Delta}_{IJ}$ are rank-$4$ differential operators
with the same general structure as \eqref{eq:hd-operator}, but different normalization of the $\nabla^4$ term. In particular
\begin{equation}\label{eq:cardy-riva-hd-operators}
 \begin{split}
 \hat{\Delta}_{\phi\phi} = (2l_1+l_2) (\nabla^2)^2 +\cdots\,, \quad
 \hat{\Delta}_{\chi\chi} = 4 l_1 (\nabla^2)^2 +\cdots\,,
 \end{split}
\end{equation}
with the dots hiding lower-derivative terms of rank-$2$,
so the method of Appendix~\ref{sect:hk} applies after appropriately rescaling the two fields.
A significant simplification comes from the fact that the two scalars have different parity, so the off-diagonal terms $\hat{\Delta}_{\phi\chi}$ and $\hat{\Delta}_{\chi\phi}$ do not contribute, which can be checked explicitly. In practice, the anomaly is the same as that of two decoupled scalars with action
\begin{equation}
 \begin{split}
 \frac{1}{2}\int {\rm d}^2 x \sqrt{g}
 \Bigl\{
 \phi \hat{\Delta}_{\phi\phi} \phi
 +\chi \hat{\Delta}_{\chi\chi} \chi
 \Bigr\}\,.
 \end{split}
\end{equation}

After integrating out $\phi$ and $\chi$ using the method sketched in Appendix~\ref{sect:hk}, a somewhat lengthy computation that uses twice the same formula as \eqref{eq:hk-computation} gives the anomaly
\begin{equation}
 \begin{split}
 A_\sigma &= \frac{1}{2\pi}\int {\rm d}^2 x \sqrt{g}\Bigl\{
 \frac{13l_1+5l_2}{6(2l_1+l_2)} R
 -\frac{3l_1+l_2}{2l_1+l_2} \nabla^\mu S_\mu
 \\& \qquad\qquad\qquad
 -\frac{(3l_1+l_2)^2}{4l_1 (2l_1+l_2)} S_\mu S^\mu
 \Bigr\}
 \,.
 \end{split}
\end{equation}
The nontrivial ``denominators'' $2l_1+l_2$ and $4l_1$
are caused by the fact that, for the application of the heat kernel methods of Appendix~\ref{sect:hk} to the differential operators \eqref{eq:cardy-riva-hd-operators}, we have to rescale the fields $\phi$ and $\chi$, respectively.

Comparing the anomaly with \eqref{eq:anomaly}, we find the following values of the charges
\begin{equation}
 \begin{split}
 &\beta_\Phi = \frac{5}{3}+\frac{l_1}{(2l_1+l_2)}
 \,,\qquad
 \beta_\Psi = \frac{2}{3}
 \,, 
 \\&
 \beta^S_2 = -\frac{(3l_1+l_2)^2}{4l_1 (2l_1+l_2)}\,.
 \end{split}
\end{equation}
The important difference from the example of the Sect.~\ref{sect:example} is that $\beta^S_2\neq 0$ \emph{unless} the global conformality condition $3l_1+l_2=0$
is satisfied.
This result is somewhat expected: in the global conformal limit the charge $\beta^S_2$, that represents a departure from the flow $\mu \frac{\rm d}{{\rm d}\mu} \beta_\Psi = \chi_{ij}\beta^i\beta^j$, is zero. Instead, when the theory is only scale invariant we have that $\beta^S_2$ is different from zero. The global conformal limit has the additional property that
$
 \beta_\Phi = \beta_\Psi = \frac{2}{3}
$. The general condition $\beta_\Phi = \beta_\Psi$ was observed as a consequence of the decoupling of $S_\mu$ for conformal theories, so it 
further confirms that the limit is globally conformally invariant.

\section{Interplay between Weyl and diffeomorphisms symmetries}\label{sect:interplay}

In this section we want to discuss some points that may be of relevance for the application of the Weyl gauged local rg to quantum gravity.
It is known that, as a result of the path-integral procedure, one can choose to either have a conformal or a diffeomorphism anomaly. Classically, the energy momentum tensor can be made traceless at the price of its conservation. For a quantum example, see Ref.~\cite{Karakhanian:1994gs} with a discussion of the case $d=2$, which becomes very relevant when studying two-dimensional quantum gravity because the Einstein action is not dynamical in the limit \cite{Martini:2021slj,Martini:2023qkp}. The presence of the Weyl anomaly discussed in the previous sections assumes tacitly that diffeomorphisms invariance is not anomalous.

The gauged Weyl transformations \eqref{eq:gauged-weyl-transformations} for infinitesimal $\sigma$ are
\begin{equation}
 \begin{split}
  &\delta^W_\sigma g_{\mu\nu} = 2 \sigma g_{\mu\nu}\,,
  \qquad
  \delta^W_\sigma S_\mu = -\partial_\mu \sigma\,,
  \\&
  \delta^W_\sigma \Phi = 2 w_\Phi \sigma \Phi\,,
 \end{split}
\end{equation}
and, as discussed in Sect.~\ref{sect:intro}, imply classically $T=\nabla^\mu D_\mu$ (the label $W$ is for ``Weyl''). The classical relation between the currents is subject to the anomaly \eqref{eq:anomaly} coming from the quantization, as we discussed in Sect.~\ref{sect:two}.

As stressed above, we have been working under the assumption that the classical action $\Gamma$ is diffeomorphism invariant. Infinitesimal diffeomorphisms are generated by a vector $\xi^\mu$
\begin{equation}
 \begin{split}
  &
  \delta^E_\xi g_{\mu\nu} = ({\cal L}_\xi g)_{\mu\nu} = 2\nabla_{(\mu} \xi_{\nu)}\,,
  \\
  &
  \delta^E_\xi S_\mu = ({\cal L}_\xi S)_\mu
  = \xi^\nu \nabla_\mu S_\nu
  +S^\nu \nabla_\mu \xi_\nu
  \,,
  \quad
  \delta^E_\xi \Phi = {\cal L}_\xi \Phi
  \,,
 \end{split}
\end{equation}
where we used the fact that $\nabla_\mu$ is symmetric and compatible (the label $E$ is for ``Einstein'').
The tensor structure of $\Phi$ is unspecified, so the Lie derivative of $\Phi$ depends on its holonomic indices, while it is insensitive to other indices, which could be gauge and local Lorentz ones, and it is also insensitive to the Weyl weight. Consequently, $\delta^E_\xi \Phi$ might not transform covariantly under the other local transformations \cite{Codello:2012sn}.
On-shell, classical invariance under diffeomorphisms implies that
\begin{equation}\label{eq:diff-invariance-consequence}
 \begin{split}
  \nabla_\mu T^{\mu\nu} + \nabla_\mu D^\mu S^\nu + D_\mu W^{\mu\nu}=0
  \,,
 \end{split}
\end{equation}
and the energy-momentum tensor is conserved in the limit $S_\mu=W_{\mu\nu}=0$.
This equation is not anomalous if diffeomorphisms invariance is preserved during quantization.

With a bit of work, it is possible to rewrite the relation \eqref{eq:diff-invariance-consequence} in a gauged Weyl-covariant form
\begin{equation}\label{eq:diff-invariance-consequence-covariant}
 \begin{split}
  \hat{\nabla}_\mu T^{\mu\nu} + D_\mu W^{\mu\nu}=0
  \,,
 \end{split}
\end{equation}
where we used the fact that $w(T^{\mu\nu})= -d-2$ (which can be deduced from its definition, given that the Weyl weight is multiplicative), but also, importantly, we used the relation 
$T=\nabla^\mu D_\mu$ coming from classical gauged Weyl invariance.
Classically, the energy-momentum tensor is thus gauge-covariantly conserved also for integrable Weyl geometries in which $W_{\mu\nu}=0$ (i.e., when $S_\mu$ is locally the gradient of some scalar function).

A more direct way to prove the gauge-covariant relation \eqref{eq:diff-invariance-consequence-covariant} is to consider ``modified'' diffeomorphism transformations, which combine diffeomorphisms and Weyl transformations as
\begin{equation}\label{eq:mod-diff}
 \begin{split}
  \tilde{\delta}^E_\xi = \delta^E_\xi + \delta^W_{\xi \cdot S}
  \,,
 \end{split}
\end{equation}
where $\xi \cdot S=\xi^\mu S_\mu$. This transformation makes it such that
$\tilde{\delta}^E_\xi {\cal O}$ is Weyl-covariant if ${\cal O}$ is also Weyl covariant, for an arbitrary operator ${\cal O}$ with some weight and holonomic indices. The operator ${\cal O}$ could be the basic field $\Phi$, or any of the ${\cal O}_i$. A generalization to other anholonomic and gauge indices is possible by including in \eqref{eq:mod-diff} further contributions of the form $\delta^{G}_{\xi\cdot A}$, where $A_\mu= t^I A^I_\mu$ is the potential of some gauge group $G$ and algebra generators $t^I$.
In practical applications the transformation $\tilde{\delta}^E_\xi$ is the so-called covariant Lie derivative \cite{Obukhov:2007se}, sometimes denoted $\tilde{\delta}^E_\xi = \tilde{{\cal L}}_\xi$ \cite{Sauro:2022chz}.

It is interesting to study the algebras, in view of the application of Wess-Zumino consistency conditions. We have for the original transformations
\begin{equation}
 \begin{split}
  [\delta^W_\sigma,\delta^W_{\sigma'}]=0
  \,,\qquad
  [\delta^E_\xi,\delta^E_{\xi'}] = \delta^E_{[\xi,\xi']}\,,
 \end{split}
\end{equation}
where $[\xi,\xi']^\mu$ are the Lie brackets. The algebra of diffeomorphisms is isomorphic to the algebra of Lie brackets, as expected. Using the definition of 
$\tilde{\delta}^E_\xi$, one can show
\begin{equation}\label{eq:mod-algebra}
 \begin{split}
  [\tilde{\delta}^E_\xi,\tilde{\delta}^E_{\xi'}] = \tilde{\delta}^E_{[\xi,\xi']} + \delta^W_{W(\xi,\xi')}\,,
 \end{split}
\end{equation}
where $W(\xi,\xi')= W_{\mu\nu}\xi^\mu \xi^{\prime\nu}$ ($W= \frac{1}{2}W_{\mu\nu}{\rm d}x^\mu \wedge {\rm d}x^\nu$ is the Weyl curvature $2$-form).
The above algebra extends correctly to sources and connections, including $S_\mu$, even if it has an affine transformation \cite{Sauro:2022chz}.
We have that $\tilde{\delta}^E_\xi$ does not form a closed subalgebra, unless for integrable Weyl geometries when $W(\xi,\xi')=0$. The algebra satisfies a Jacobi identity, but it does not generate a group in the traditional sense, because it depends on $S_\mu$ through $W_{\mu\nu}$.

The relation \eqref{eq:mod-algebra}
makes it clear that if one chooses to preserve
diffeomorphisms invariance (i.e., one assumes that diffeomorphisms invariance is preserved by the underlying path-integral's measure,
but Weyl transformation are anomalous),
then the modified diffeomorphisms are anomalous, because of the presence of the anomaly in $\delta^W_{W(\xi,\xi')}$. We deduce the quantum version of \eqref{eq:diff-invariance-consequence-covariant} using the anomaly
\begin{equation}
 \begin{split}
  \langle \hat{\nabla}_\mu T^{\mu\nu}\rangle + \langle D_\mu \rangle W^{\mu\nu} + \langle S^\nu {\cal A}\rangle =0
 \end{split}
\end{equation}
where ${\cal A}= \frac{\delta}{\delta \sigma}(\Delta^\beta_\sigma \Gamma+ A_\sigma)$ is the anomalous part of the trace of the energy-momentum tensor, using the definitions give in Eqs.~\eqref{eq:anomaly} and \eqref{eq:beta-delta}. Therefore, even at a fixed point $\beta^i=0$ and even for integrable Weyl geometries, we expect an anomalous contribution to the gauge-covariant conservation of the energy-momentum tensor.
Using the explicit form of the anomaly, we can inspect the limit $\beta_i=0$
for constant (traditional) couplings $\lambda^i$, we find
\begin{equation}
 \begin{split}
  &\langle \hat{\nabla}_\mu T^{\mu\nu}\rangle + \langle D_\mu \rangle W^{\mu\nu} + \frac{\beta_\Phi}{4\pi}\hat{R} S^\nu
  \\&
  \quad +\frac{\tilde{\beta}_\Psi}{2\pi}\nabla^\mu S_\mu S^\nu
  +\frac{\beta^S_2}{2\pi} S^\mu S_\mu S^\nu
  =0
  \,,
 \end{split}
\end{equation}
which depends, as one might have expected, on the charge $\tilde{\beta}_\Psi$
that can be monotonic if $\beta^S_2=0$ and $\chi_{ij}>0$ (the scalar functions of the couplings are here regarded as constant numbers set at their fixed point values). We are treating the sources as classical fields, so further expectation values, for example those including insertions of ${\cal O}_i$, could be obtained using functional derivatives of the local couplings.

\section{Conclusions} \label{sect:conclusions}

The generalization of the local rg approach to theories involving the gauged version of Weyl transformations reveals interesting similarities, but also several differences, when compared to the nongauged case developed in the past \cite{Osborn:1991gm}.
In practice, the generalization consists to coupling the finite and renormalized quantum theory with local couplings to a Weylian geometry, rather than a Riemannian one. The former is characterized by the presence of an additional gauge potential $S_\mu$ that acts as a souce of the dilation current and that has an affine transformation under local rescalings of the metric. The dilation current is naturally interpreted as the virial current of scale invariant models in the flat space limit.

We have restricted our attention to the application of the Wess-Zumino consistency condition to the anomaly in the two dimensional case, although several considerations can be straightforwardly extended to four or higher dimensions.
Our results indicate that a new charge, denoted $\beta^S_2$ in the main text, is responsible for the differences between scale and conformal invariance.
While we see that it is not possible to set the value of the new charge to zero
in an arbitrary renormalization scheme of a general model, we also realize that, if for some specific theory $\beta^S_2=0$, then it should be possible
to find consistent rg trajectories such that the new charge remains zero along the flow.
However, we do not expect a generalization of Zamolodchikov's irreversibility theorem \cite{Zamolodchikov:1986gt} on such consistent trajectories: the theorem requires the positivity of the metric in the space of the couplings, $\chi_{ij}>0$, which in turn requires the underlying model to be unitary \cite{Nakayama:2013is}, but, to our knowledge, all examples of scale-but-not-conformal theories are nonunitary \cite{Gimenez-Grau:2023lpz}.

The natural extension of our work is to repeat the analysis in the four dimensional case, which is certainly going to be much more complicate
since we already know that the four dimensional local rg requires the inclusion of several more tensor structures in the space of the marginal couplings \cite{Osborn:1991gm}. Nevertheless, we think that the four dimensional analysis is certainly within reach. The discussion of the four dimensional case would also lead to more interesting example applications
as compared to the ones shown in the present paper. The four dimensional case could also be important for the study of completions of  ``Standard Model $+$ Gravity'' that require gauged Weyl symmetry \cite{Ghilencea:2023wwf}. For these completions of the Standard Model it has been shown that the Weyl anomaly cancels (for standard constant couplings), also thanks to the fact that $S_\mu$ couples only to the Higgs sector \cite{Ghilencea:2021lpa}. The four dimensional anomaly with local couplings that generalizes \eqref{eq:anomaly} can nevertheless offer nonperturbative insights on the rg flow.

\smallskip

\paragraph*{Acknowledgments.} The author is grateful to H.~Gies, G.~Paci and D.~Sauro for discussions and comments that influenced the development of this work. The author is also grateful to an anonymous referee for correcting several statements on irreversibility.

\appendix

\section{Riemann and Weyl geometries} \label{sect:geometry}

Geometrically, a Riemannian geometry consists of a pair $(M,g_{\mu\nu})$, where $M$ is a manifold and $g_{\mu\nu}$ a metric. In contrast, a Weylian geometry can be seen as a triple $(M,g_{\mu\nu},S_\mu)$ such that the geometrical information depends on the equivalence $(g_{\mu\nu},S_\mu) \sim ({\rm e}^{2\sigma}g_{\mu\nu},S_\mu-\partial_\mu \sigma)$.

We can associate local curvatures to the connection $\hat{\nabla}$ defined in \eqref{eq:nabla-hat} \cite{Sauro:2022chz}.
The simplest curvature is the gauge invariant field strength, $W_{\mu\nu}=\partial_\mu S_\nu-\partial_\nu S_\mu$.
To work out the others, consider a holonomic vector field $v^\mu$ with zero Weyl weight $w(v^\mu)=0$.
We have the commutator
\begin{equation}
 \begin{split}
  [\hat{\nabla}_\mu,\hat{\nabla}_\nu]v^\alpha = \hat{R}^\alpha{}_{\beta\mu\nu} v^\beta
 \end{split}
\end{equation}
and $\hat{R}^\alpha{}_{\beta\mu\nu}= R^\alpha{}_{\beta\mu\nu}+\cdots$, where the dots hide several additional structures that depend on $S_\mu$ and its covariant derivatives, which we do not report for brevity.

A nonzero weight of $v^\mu$
would just give additional structures involving $W_{\mu\nu}$,
so what follows is only one possible convention to define the Riemann and Ricci tensors of $\hat{\nabla}$ \cite{Sauro:2022chz}. It is also important to realize that the ``charge'' of Weyl transformation, i.e., the weight, is multiplicative, so it depends on whether the indies are raised or lowered with the metric. For example $v_\mu$ weighs differently than $v^\mu$. In our convention, $w(v_\mu)= w(g_{\mu\nu})+w(v^\mu) = w_g= 2 \neq 0$ if $w(v^\mu)=0$.

We can define a Ricci tensor as $\hat{R}_{\mu\nu} = \hat{R}^\alpha{}_{\mu\alpha\nu}$, so
\begin{equation}
 \begin{split}
  \hat{R}_{\mu\nu} &= 
  R_{\mu\nu} - g_{\mu\nu} \nabla^\alpha S_\alpha
  \\&- (d-2)\Bigl\{
  \nabla_{(\mu} S_{\nu)} -S_\mu S_\nu + g_{\mu\nu} S^\alpha S_\alpha
  \Bigr\}
  -\frac{d}{2} W_{\mu\nu}\,,
 \end{split}
\end{equation}
which is not symmetric due to the presence of $W_{\mu\nu}$ (as seen from the last term). The contraction leading to $\hat{R}_{\mu\nu}$ is not unique, but alternative choices differ by the above only by the coefficient of the antisymmetric part. Consequently, the Ricci scalar $\hat{R}=g^{\mu\nu}\hat{R}_{\mu\nu}$ is unambiguous,
\begin{equation}
 \begin{split}
  \hat{R}= R -2(d-1)\nabla^\mu S_\mu -(d-1)(d-2) S^\mu S_\mu\,.
 \end{split}
\end{equation}
Let us also note that the gauged Weyl geometry admits an ambient space construction \cite{Jia:2023gmk}, in analogy to the one of conformal geometry \cite{Fefferman:2007rka}.

\section{Heat kernel}\label{sect:hk}

The heat kernel of an elliptic differential operator $\hat{\Delta}$ is defined as the solution of the diffusion equation
\begin{equation}
 \begin{split}
  (\partial_t + \hat{\Delta}_x ) {\cal H}_{\hat{\Delta}}(s;x,x') =0\,,
  \end{split}
\end{equation}
with initial condition ${\cal H}_{\hat{\Delta}}(0;x,x')=\delta(x,x')$, where $\delta(x,x')$ is the covariant Dirac delta \cite{Vassilevich:2003xt}.
Formally, we have the representation
${\cal H}_{\hat{\Delta}}(t;x,x')=\langle x|\exp(-t \hat{\Delta})| x' \rangle$. For the computations of Sect~\ref{sect:examples}, $\hat{\Delta}$ is a rank-$2p$ scalar operator of the form $\hat{\Delta} = (-\nabla^2)^p +\cdots$, where the dots include lower order terms in the covariant derivatives, and
we only need the coincidence limit $x=x'$ of the heat kernel, which admits an asymptotic expansion in $s$
\begin{equation}
 \begin{split}
  {\cal H}_{\hat{\Delta}}(s;x,x) =
  \frac{1}{(4\pi)^{d/2}t^{1/p}}\left(1+\sum_{n\geq 1} t^{n/p} a_n(x)\right)
  \,.
  \end{split}
\end{equation}
The coefficients $a_n(x)$ are local functions depending on metric, potentials, curvatures and tensor structures of $\hat{\Delta}$. The $a_n(x)$ can be computed with various algorithms \cite{Vassilevich:2003xt} and are known also for the general case $p=2$ \cite{Barvinsky:2021ijq}, which includes the operator \eqref{eq:hd-operator} given in Sect.~\ref{sect:example}. Similar operators are needed in Sect.~\ref{sect:cardy-riva}.

The heat kernel is directly related to the zeta-function of the differential operator by
$\zeta_{\hat{\Delta}}(s;x,x') =
{\Gamma(s)}^{-1} \int_{0}^{\infty} {\rm d}t \, t^{s-1}  {\cal H}_{\hat{\Delta}}(t;x,x')$
and the asymptotic expansion is used to evaluate $\zeta_{\hat{\Delta}}(0,x,x)=(4\pi)^{-d/2} a_{d/2}(x)$. Using the zeta-function and considering a bare action of the form $S\sim \int \varphi \hat{\Delta} \varphi$, we can formally express the nonrenormalized effective action as
\begin{equation}
 \begin{split}
  \Gamma = -\frac{1}{2} {\rm Tr} \left. \frac{{\rm d}}{{\rm d} s} \zeta_{\hat{\Delta}}(s;x,x') \right|_{s=0}\,,
  \end{split}
\end{equation}
where the trace includes an integration over spacetime for $x=x'$.
For Weyl and diffeomorphism covariant differential operators, such as the ones
considered in the main text in Sect.~\ref{sect:examples}, it is straightforward to show that the scale transformation of the renormalized action is
\begin{equation}
 \begin{split}
  \delta_\sigma \Gamma &=
  \frac{1}{(4\pi)^{d/2}} \int {\rm d}^d x \sqrt{g} \, \sigma a_{d/2}(x)\,,
  \end{split}
\end{equation}
and therefore we have a formal way to obtain the anomaly.
From the above computation follows the standard result that shows the relation of the anomaly with the coefficient of the dimensional pole of divergent part of the regulated effective action.
The case $d=2$ requires the heat kernel coefficient $a_1(x)$, which, for a self-adjoint differential operator of rank four acting on a scalar,
$\hat{\Delta}=(\nabla^2)^2+B^{\mu\nu} \nabla_\mu \partial_\nu
+C^\mu \partial_\nu +D $ such as \eqref{eq:hd-operator}, is given in completely general form in Ref.~\cite{Barvinsky:2021ijq}
\begin{equation}
 \begin{split}
  a_1(x) &= \frac{\Gamma(d/4-1/2)}{\Gamma(d/2-1)}\Bigl(
  \frac{1}{2d} B^\mu{}_\mu +\frac{1}{6} R
  \Bigr)\,.
  \end{split}
\end{equation}
Notice that heat kernel coefficients of differential operators of rank greater than two depend on $d$ nontrivially, so it is necessary to take the limit $d=2$, which appears in Eq.~\eqref{eq:hk-computation} of the main text.



\begin{thebibliography}{99}

\bibitem{Shore:1986hk}
G.~M.~Shore,
Nucl. Phys. B \textbf{286}, 349-377 (1987)

\bibitem{Wess:1971yu}
J.~Wess and B.~Zumino,
Phys. Lett. B \textbf{37}, 95-97 (1971)

\bibitem{Osborn:1991gm}
H.~Osborn,
Nucl. Phys. B \textbf{363}, 486-526 (1991)

\bibitem{Zamolodchikov:1986gt}
A.~B.~Zamolodchikov,
JETP Lett. \textbf{43}, 730-732 (1986)

\bibitem{Jack:1990eb}
I.~Jack and H.~Osborn,
Nucl. Phys. B \textbf{343}, 647-688 (1990)

\bibitem{Cardy:1988cwa}
J.~L.~Cardy,
Phys. Lett. B \textbf{215}, 749-752 (1988)

\bibitem{Iorio:1996ad}
A.~Iorio, L.~O'Raifeartaigh, I.~Sachs and C.~Wiesendanger,
Nucl. Phys. B \textbf{495}, 433-450 (1997)
[arXiv:hep-th/9607110 [hep-th]].

\bibitem{Nakayama:2013is}
Y.~Nakayama,
Phys. Rept. \textbf{569}, 1-93 (2015)
[arXiv:1302.0884 [hep-th]].

\bibitem{Codello:2012sn}
A.~Codello, G.~D'Odorico, C.~Pagani and R.~Percacci,
Class. Quant. Grav. \textbf{30}, 115015 (2013)
[arXiv:1210.3284 [hep-th]].

\bibitem{Sauro:2022chz}
D.~Sauro and O.~Zanusso,
Class. Quant. Grav. \textbf{39}, no.18, 185001 (2022)
[arXiv:2203.08692 [hep-th]].

\bibitem{Brown:1980qq}
L.~S.~Brown and J.~C.~Collins,
Annals Phys. \textbf{130}, 215 (1980)

\bibitem{OsbornLectures}
H.~Osborn,
\textit{``{Lectures on Conformal Field Theories in more than two
  dimensions}''},
\href{http://www.damtp.cam.ac.uk/user/ho/CFTNotes.pdf}{\texttt{http://www.damtp.cam.ac.uk/user/ho/CFTNotes.pdf}}.

\bibitem{Polchinski:1987dy}
J.~Polchinski,
Nucl. Phys. B \textbf{303}, 226-236 (1988)

\bibitem{Fefferman:2007rka}
C.~Fefferman and C.~R.~Graham,
Ann. Math. Stud. \textbf{178}, 1-128 (2011)
[arXiv:0710.0919 [math.DG]].

\bibitem{Weyl:1918ib}
H.~Weyl,
Sitzungsber. Preuss. Akad. Wiss. Berlin (Math. Phys. ) \textbf{1918}, 465 (1918)

\bibitem{Sauro:2022hoh}
D.~Sauro, R.~Martini and O.~Zanusso,
[arXiv:2208.10872 [hep-th]].

\bibitem{Karananas:2015eha}
G.~K.~Karananas and A.~Monin,
Phys. Rev. D \textbf{93}, no.6, 064013 (2016)
[arXiv:1510.07589 [hep-th]].

\bibitem{Tomboulis:2011qh}
E.~T.~Tomboulis,
Phys. Rev. D \textbf{84}, 084018 (2011)
[arXiv:1105.5848 [hep-th]].

\bibitem{Ghilencea:2023wwf}
D.~M.~Ghilencea and C.~T.~Hill,
[arXiv:2303.02515 [hep-th]].

\bibitem{Fortin:2012hn}
J.~F.~Fortin, B.~Grinstein and A.~Stergiou,
JHEP \textbf{01}, 184 (2013)
[arXiv:1208.3674 [hep-th]].

\bibitem{Morris:2018zgy}
T.~R.~Morris and R.~Percacci,
Phys. Rev. D \textbf{99}, no.10, 105007 (2019)
[arXiv:1810.09824 [hep-th]].

\bibitem{Andriolo:2022lcb}
E.~Andriolo, V.~Niarchos, C.~Papageorgakis and E.~Pomoni,
Phys. Rev. D \textbf{107}, no.2, 025006 (2023)
[arXiv:2210.10891 [hep-th]].

\bibitem{Jack:2013sha}
I.~Jack and H.~Osborn,
Nucl. Phys. B \textbf{883}, 425-500 (2014)
[arXiv:1312.0428 [hep-th]].

\bibitem{Deser:1993yx}
S.~Deser and A.~Schwimmer,
Phys. Lett. B \textbf{309}, 279-284 (1993)
[arXiv:hep-th/9302047 [hep-th]].

\bibitem{Brust:2016gjy}
C.~Brust and K.~Hinterbichler,
JHEP \textbf{02}, 066 (2017)
[arXiv:1607.07439 [hep-th]].

\bibitem{Riva:2005gd}
V.~Riva and J.~L.~Cardy,
Phys. Lett. B \textbf{622}, 339-342 (2005)
[arXiv:hep-th/0504197 [hep-th]].

\bibitem{Martini:2023rnv}
R.~Martini, G.~Paci and D.~Sauro,
[arXiv:2304.08360 [gr-qc]].

\bibitem{Vassilevich:2003xt}
D.~V.~Vassilevich,
Phys. Rept. \textbf{388}, 279-360 (2003)
[arXiv:hep-th/0306138 [hep-th]].

\bibitem{Barvinsky:2021ijq}
A.~O.~Barvinsky and W.~Wachowski,
Phys. Rev. D \textbf{105}, no.6, 065013 (2022)
[arXiv:2112.03062 [hep-th]].

\bibitem{Fradkin:1982xc}
E.~S.~Fradkin and A.~A.~Tseytlin,
Phys. Lett. B \textbf{110}, 117-122 (1982)
[erratum: Phys. Lett. B \textbf{126}, 506 (1983)]

\bibitem{Riegert:1984kt}
R.~J.~Riegert,
Phys. Lett. B \textbf{134}, 56-60 (1984)

\bibitem{Paneitz}
S.~M.~Paneitz, 2008, SIGMA, 4, 036.

\bibitem{Paci:2023twc}
G.~Paci, D.~Sauro and O.~Zanusso,
Class. Quant. Grav. \textbf{40}, no.21, 215005 (2023)
[arXiv:2302.14093 [hep-th]].

\bibitem{AF}
M.~E.~Fisher and A.~Aharony
Phys. Rev. Lett. 30, 559 (1973).

\bibitem{Gimenez-Grau:2023lpz}
A.~Gimenez-Grau, Y.~Nakayama and S.~Rychkov,
[arXiv:2309.02514 [hep-th]].

\bibitem{Mauri:2021ili}
A.~Mauri and M.~I.~Katsnelson,
Nucl. Phys. B \textbf{969}, 115482 (2021)
[arXiv:2104.06859 [cond-mat.stat-mech]].

\bibitem{Zanusso:2014wsa}
O.~Zanusso,
Phys. Rev. E \textbf{90}, no.5, 052110 (2014)
[arXiv:1408.0741 [cond-mat.stat-mech]].

\bibitem{Nakayama:2016dby}
Y.~Nakayama,
Annals Phys. \textbf{372}, 392-396 (2016)
[arXiv:1604.00810 [hep-th]].

\bibitem{Karakhanian:1994gs}
D.~R.~Karakhanian, R.~P.~Manvelyan and R.~L.~Mkrtchian,
Phys. Lett. B \textbf{329}, 185-188 (1994)
[arXiv:hep-th/9401031 [hep-th]].

\bibitem{Martini:2021slj}
R.~Martini, A.~Ugolotti, F.~Del Porro and O.~Zanusso,
Eur. Phys. J. C \textbf{81}, no.10, 916 (2021)
[arXiv:2103.12421 [hep-th]].

\bibitem{Martini:2023qkp}
R.~Martini, D.~Sauro and O.~Zanusso,
[arXiv:2302.14804 [hep-th]].

\bibitem{Obukhov:2007se}
Y.~N.~Obukhov and G.~F.~Rubilar,
Phys. Rev. D \textbf{76}, 124030 (2007)
[arXiv:0712.3547 [hep-th]].

\bibitem{Ghilencea:2021lpa}
D.~M.~Ghilencea,
Eur. Phys. J. C \textbf{82}, no.1, 23 (2022)
[arXiv:2104.15118 [hep-ph]].

\bibitem{Jia:2023gmk}
W.~Jia, M.~Karydas and R.~G.~Leigh,
Nucl. Phys. B \textbf{991}, 116224 (2023)
[arXiv:2301.06628 [hep-th]].

\bibitem{Manvelyan:2007tk}
R.~Manvelyan, K.~Mkrtchyan and R.~Mkrtchyan,
Phys. Lett. B \textbf{657}, 112-119 (2007)
[arXiv:0707.1737 [hep-th]].

\end{thebibliography}
\end{document}